\documentstyle[amssymb,prl,aps,manuscript]{revtex}

\draft
\input{diagrams.tex}

\begin{document}
\author{S. Grillo \ \ \&\ \ \ H. Montani}
\address{{\it Centro At\'{o}mico Bariloche and Instituto Balseiro}\\
{\it 8400-S. C. de Bariloche}\\
Rio Negro, {\it Argentina}}
\title{Integrable mixing of $A_{n-1}$ type vertex models}
\date{November 2002}
\maketitle

\begin{abstract}
Given a family of monodromy matrices ${\sf T}_{0},{\sf T}_{1},...,{\sf T}%
_{K-1}$ corresponding to integrable anisotropic vertex models of $A_{n_{\mu
}-1}$ type, $\mu =0,1,...,K-1$, we build up a related mixed vertex model by
means of glueing the lattices on which they are defined, in such a way that
integrability property is preserved. The glueing process is implemented
through 1-dimensional representations of rectangular quantum matrix algebras 
${\rm A}\left( R_{n_{\mu -1}}:R_{n_{\mu }}\right) $, namely, the {\em %
glueing matrices }$\zeta _{\mu }$. Algebraic Bethe ansatz is applied on a
pseudovacuum space with a selected basis and, for each elements of this
basis, it yields a set of nested Bethe ansatz equations matching up to the
ones corresponding to an $A_{m-1}$ quasi-periodic model, with $m$ equal to $%
min_{\mu \in {\Bbb {Z}}_{K}}\left\{ 
\mathop{\rm rank}%
\zeta _{\mu }\right\} $. \newline
\end{abstract}


\section{Introduction}

There exists a deep linking between solvable two-dimensional vertex models
in statistical mechanics and the quantum Yang-Baxter (YB) equation, where it
appears as condition for integrability related to some basic quantities of
the model \cite{baxter}\cite{faddeev}. This linking relies on the existence
of an underlying symmetry, the quantum group \cite{d}\cite{frt}, which comes
to provide a nice algebraic framework to study these systems. There,
solutions of the YB equation are representations of some quadratic relations
defining the quantum group structure and, through a process called
baxterization \cite{jones}, they connect with the monodromy matrices of the
model. From an algebraic point of view, baxterization makes an ordinary
quantum group into a YB algebra. In this way, different representations of a
YB algebra lead to different integrable models.

The main aim of this work is to present a glueing process of models
associated to several YB algebras preserving the integrability of the total
system. We restrict ourself to those YB algebras ${\rm YB}_{n}$ coming from
baxterization of the quantum groups ${\rm A}\left( R_{n}\right) $ \cite{frt}%
, i.e., the duals of $U_{q}\left( su_{n}\right) $ \cite{d}, $n\in {\Bbb N}$.
Beside algebras ${\rm YB}_{n}$, the glueing process also involves their
rectangular generalizations ${\rm YB}_{n,m}$, defined as the spectral
parameter dependent versions of the rectangular quantum matrix algebras $%
{\rm A}\left( R_{n}:R_{m}\right) $ \cite{mm}. Here ${\rm A}\left(
R_{n}:R_{n}\right) ={\rm A}\left( R_{n}\right) $ and, accordingly, ${\rm YB}%
_{n,n}={\rm YB}_{n}$. The process is based on the existence of algebra
homomorphisms ${\rm YB}_{n,m}\rightarrow {\rm YB}_{n,p}\otimes {\rm YB}%
_{p,m} $, the cocomposition maps, that generalize the concept of coproduct
in a bialgebra. Such maps can be used to build up representations of a given
YB algebra ${\rm YB}_{n}$ as a product of representations of another
algebras ${\rm YB}_{m}$, $m\neq n$, in an analogous way as the standard
coproduct is used for building up usual tensor representations. More
precisely, given families ${\sf T}_{\mu }$ and $\zeta _{\mu }$, $\mu
=0,1,...,K-1$, of representations of ${\rm YB}_{n_{\mu }}$ and ${\rm YB}%
_{n_{\mu -1},n_{\mu }}$, respectively, cocomposition maps ensure operator $%
{\sf T}^{mix}=\zeta _{0}\stackrel{.}{\otimes }{\sf T}_{0}\stackrel{.}{%
\otimes }...\stackrel{.}{\otimes }\zeta _{K-1}\stackrel{.}{\otimes }{\sf T}%
_{K-1}$ is a representation of ${\rm YB}_{n_{K-1}}$ (symbol $\stackrel{.}{%
\otimes }$ will be defined in next section). If each ${\sf T}_{\mu }$
defines the monodromy matrix of a given vertex model, we say ${\sf T}^{mix}$
is that of the mixed model with glueing matrices $\zeta _{\mu }$.

We shall see this procedure is compatible with the algebraic Bethe ansatz
method for solving these models, in the sense there exists a set of
pseudovacuum vectors with respect to which these technics can be applied.
Moreover, we show a set of nested Bethe ansatz equations identical to the
ones corresponding to an $A_{m-1}$ quasi-periodic model, with $m$ equal to $%
min_{\mu \in {\Bbb {Z}}_{K}}\left\{ 
\mathop{\rm rank}%
\zeta _{\mu }\right\} $, is related to each one of these vectors. \newline

This work is organized as follows: in section {\bf II}, we review some well
known facts on the connection between YB algebras and integrable vertex
models; in section {\bf III}, we describe the glueing process and the
glueing matrices as 1-dimensional representations of the rectangular YB
algebras; finally in section {\bf IV}, we prove complete integrability of
mixed models, showing that diagonalization of mixed transfer matrices reduce
to solve nested Bethe equations of a family of $A$-type vertex models.

\section{Yang-Baxter algebras and integrable vertex models}

To start with, we describe briefly the connection between two-dimensional
vertex models and $A_{n-1}$ type solutions of the YB equation.

Let us consider the class of YB operators or constant $R$-matrices 
\begin{equation}
\left[ R_{n}\right] _{ab}^{kl}=\left\{ 
\begin{array}{ll}
q\,\delta _{a}^{k}\,\delta ^{kl}, & a=b; \\ 
\delta _{a}^{k}\,\delta _{b}^{l}+\left( q-1/q\right) \,\delta
_{a}^{l}\,\,\delta _{b}^{k}, & a<b; \\ 
\delta _{a}^{k}\,\delta _{b}^{l}, & a>b;
\end{array}
\right. \;\;\;1\leq a,b,k,l\leq n;  \label{r}
\end{equation}
$q\in {\Bbb C}\backslash \left\{ 0,1\right\} $, related to the standard Hopf
algebra deformations of the simple Lie algebras $A_{n-1}$, i.e., the quantum
groups $U_{q}\left( su_{n}\right) $ and ${\rm A}\left( R_{n}\right) $, $n\in 
{\Bbb N}$. Baxterization process yields the spectral parameter dependent
versions $R_{n}\left( x\right) =xR_{n}-PR_{n}^{-1}P/x$ of each $R_{n}$, with 
$P_{ij}^{kl}=\delta _{i}^{l}\delta _{j}^{k}$ the permutation matrix and $%
x\in {\Bbb C}$. Then, for every $N\in {\Bbb N}$ a related integrable
(inhomogeneous) lattice model \cite{dev} is defined by a monodromy matrix $%
{\sf T}\doteq {\sf T}^{\left( n,N\right) }\left( x;{\bf \alpha }\right) $
with entries (sum over repeated indices convention is assumed) 
\begin{equation}
{\sf T}_{a}^{b}={\sf R}_{a}^{b_{1}}\left( \left. x\right/ \alpha _{0}\right)
\otimes {\sf R}_{b_{1}}^{b_{2}}\left( \left. x\right/ \alpha _{1}\right)
\otimes ...\otimes {\sf R}_{b_{N-1}}^{b}\left( \left. x\right/ \alpha
_{N-1}\right) ,\;\;1\leq a,b\leq n,  \label{t}
\end{equation}
being ${\bf \alpha }=\left( \alpha _{0},...,\alpha _{N-1}\right) $ a vector
of ${\Bbb C}^{N}$. Operators ${\sf R}_{a}^{b}\left( x\right) :{\Bbb C}%
^{n}\rightarrow {\Bbb C}^{n}$ are entries of a matrix ${\sf R}={\sf R}\left(
x\right) $ such that $\left[ {\sf R}_{a}^{b}\left( x\right) \right] _{i}^{j}=%
\left[ R_{n}\left( x\right) \right] _{ai}^{bj}$ in the canonical basis of $%
{\Bbb C}^{n}$. Compact notation ${\sf T}={\sf R}\stackrel{.}{\otimes }...%
\stackrel{.}{\otimes }{\sf R}$, where $\stackrel{.}{\otimes }$ denotes
matrix multiplication between consecutive factors, will be used. Note that $%
{\sf T}^{\left( n,1\right) }={\sf R}^{n}={\sf R}$. These models are the
anisotropic analogous of the $A_{n-1}$ invariant vertex models with periodic
boundary conditions. Quasi-periodic versions (see \cite{dev} again) are
given by elements $\Upsilon $ in the symmetry group of $R_{n}\left( x\right) 
$, i.e., $\Upsilon \in GL\left( n\right) $ and $\left[ R_{n}\left( x\right)
,\Upsilon \otimes \Upsilon \right] =0$. Related monodromy matrices read $%
{\sf T}^{\Upsilon }={\sf T}\cdot \Upsilon $. Equation $\left( \ref{t}\right) 
$ defines operators ${\sf T}_{a}^{b}:\left( {\Bbb C}^{n}\right) ^{\otimes
N}\rightarrow \left( {\Bbb C}^{n}\right) ^{\otimes N}$ that describe the
statistical weights assigned to each vertex configuration in a given row of
the lattice, graphically, 
\[
\begin{array}{c}
\left[ {\sf T}_{a}^{b}\right] _{i_{0},...,i_{N-1}}^{j_{0},...,j_{N-1}}=%
{\small 
\begin{diagram}[size=1.5em,height=1.3em,shortfall=0em,midshaft]
& &{} & &{} & & & &{} & & \\
 & & \uLine^{j_{0}} & & \uLine_{j_{1}} & & & & \uLine^{j_{N-1}} & & \\
 \;&\rLine_{a} & \bullet & \rLine^{\;b_1\;}& \bullet & \rLine_{\;\;b_2\;\;}&\; ...\;\;&\rLine_{b_{N-1}} & \bullet & \rLine^{b}&{} \\
 & & \uLine_{i_{0}} & &\uLine^{i_{1}} & & & & \uLine_{i_{N-1}} & & \\ & &{} & &{} & & & &{} & & \\
 \end{diagram}%
%
}\;{\small ;\;\;\;\;}\left[ R_{n}\left( x\right) \right] _{ai}^{bj}={\small 
\begin{diagram}[size=1.5em,height=1.3em,shortfall=0em,midshaft]
 & & & & \\
 & & \uLine^j & & \\
 & \rLine_a& \bullet & \rLine^b & \\
 & & \uLine_i & & \\
 & & & & \\
 \end{diagram}%
%
}\;.
\end{array}
\]
If lattice has $N^{\prime }$ rows, the partition function is ${\sf Z}=%
\mathop{\rm trace}%
\left( {\sf t}^{N^{\prime }}\right) $, being ${\sf t}=\sum_{a}{\sf T}%
_{a}^{a} $ the transfer matrix. On the other hand, the operators ${\sf T}%
_{a}^{b}\left( x;{\bf \alpha }\right) $, as it is well known, give a
representation of the YB algebra related to $R_{n}\left( x\right) $. This
algebra, which we shall indicate ${\rm YB}_{n}$, is generated by
indeterminates ${\sf T}_{i}^{j}\left( x\right) $, $1\leq i,j\leq n$; $x\in 
{\Bbb C}$, subject to relations 
\begin{equation}
\left[ R_{n}\left( x/y\right) \right] _{ij}^{kl}\,\noindent \noindent {\sf T}%
_{k}^{r}\left( x\right) \,\noindent \noindent {\sf T}_{l}^{s}\left( y\right)
={\sf T}_{j}^{l}\,\noindent \noindent \left( y\right) \,{\sf T}%
_{i}^{k}\left( x\right) \,\noindent \noindent \left[ R_{n}\left( x/y\right) 
\right] _{kl}^{rs};\;\;1\leq i,j,r,s\leq n.  \label{fcr}
\end{equation}
These relations entail the formal integrability of the system. In fact, by
taking the trace, one gets $\left[ {\sf t}\left( x\right) ,{\sf t}\left(
y\right) \right] =0$ for all $x,y\in {\Bbb C}$, i.e., the transfer matrix is
a generating function of {\em conserved quantities}. Beside this, the model
is effectively solved by means of algebraic Bethe ansatz (see \cite{fad}\cite
{marti}\cite{kor} and references therein), where the central ingredient is
the existence of an eigenstate $\omega \in \left( {\Bbb C}^{n}\right)
^{\otimes N}$ of each entry ${\sf T}_{a}^{a}$ (and consequently of the
transfer matrix ${\sf t}$), such that ${\sf T}_{a}^{b}\,\omega =0$ for all $%
a\neq b$ and $a\geq 2$, and ${\sf T}_{1}^{b}\,\omega \neq c\,\omega $, $%
\forall c\in {\Bbb C}$, for all $b\geq 2$. For latter convenience, let us
express ${\sf T}$ in the block form 
\[
{\sf T}=\left[ 
\begin{array}{cc}
{\sf A} & {\sf B}_{j} \\ 
{\sf C}_{i} & {\sf D}_{ij}
\end{array}
\right] ;\;\;\;1\leq i,j\leq n-1, 
\]
i.e., define ${\sf A}\doteq {\sf T}_{1}^{1}$, ${\sf B}_{j}\doteq {\sf T}%
_{1}^{j+1}$, ${\sf C}_{i}\doteq {\sf T}_{i+1}^{1}$ and ${\sf D}_{ij}\doteq 
{\sf T}_{i+1}^{j+1}$. Then, $\omega $ is an eigenstate of ${\sf A}$ and of
each diagonal entry ${\sf D}_{ii}$, fulfilling ${\sf C}_{i}\,\omega =0$ and $%
{\sf D}_{ij}\,\omega =0$ for $i\neq j$. A vector satisfying these conditions
is called {\em pseudovacuum vector}. On the other hand, since ${\sf B}%
_{j}\,\omega ={\sf T}_{1}^{j+1}\,\omega \neq c\,\omega $ for all $j$, each $%
{\sf B}_{j}\left( x\right) $ plays the role of a creation operator. Applying
them repeatedly on $\omega $ (varying $j$ from $1$ to $n-1$ and $x$
satisfying the so-called Bethe equations) we generate new eigenstates for
the transfer matrix, namely the {\em Bethe vectors}, giving {\em a priori} a
complete set of eigenstates for ${\sf t}$. In such a case we say the system
is exactly solvable or completely integrable (see \cite{baxter2} and refs.
therein). Nevertheless, sometimes not only a vector but a {\em pseudovacuum
subspace} (cf. \cite{kr}\cite{dk}) is needed in order to insure complete
integrability. This will be our case.

\bigskip

Since each ${\rm YB}_{n}$ is a bialgebra, with coalgebra structure 
\begin{equation}
\Delta :{\sf T}_{i}^{j}\left( x\right) \mapsto {\sf T}_{i}^{k}\left(
x\right) \otimes {\sf T}_{k}^{j}\left( x\right) ,\;\;\;\varepsilon :{\sf T}%
_{i}^{j}\left( x\right) \mapsto \delta _{i}^{j},  \label{cs}
\end{equation}
for every couple of monodromy matrices ${\sf T}^{\left( n,N\right) }$ and $%
{\sf T}^{\left( n,P\right) }$ as above we have another one, 
\[
{\sf T}^{\left( n,N+P\right) }={\sf T}^{\left( n,N\right) }\stackrel{.}{%
\otimes }{\sf T}^{\left( n,P\right) },\;\;with\;entries\;\;{\sf T}%
_{a}^{b\;\left( n,N+P\right) }={\sf T}_{a}^{c\;\left( n,N\right) }\otimes 
{\sf T}_{c}^{b\;\left( n,P\right) }, 
\]
giving again a representation of ${\rm YB}_{n}$. Furthermore, if $\omega $
and $\phi $ are the pseudovacuums of ${\sf T}^{\left( n,N\right) }$ and $%
{\sf T}^{\left( n,P\right) }$, then $\omega \otimes \phi $ defines a
pseudovacuum for ${\sf T}^{\left( n,N+P\right) }$. Consequently the enlarged
model, or the {\em glueing} of ${\sf T}^{\left( n,N\right) }$ and ${\sf T}%
^{\left( n,P\right) }$, is also integrable. In particular, thermodynamic
limit $N\rightarrow \infty $ preserves integrability. But, can we glue
models which give representations of different YB algebras, e.g., ${\rm YB}%
_{n}$ and ${\rm YB}_{m}$ with $n\neq m$, and such that a pseudovacuum exists
for the resulting model? The aim of this paper is to answer last question.
More precisely, we build up from a family $\left\{ {\sf T}_{\mu }:\mu \in 
{\Bbb Z}_{K}\right\} $ of {\em pure }models, i.e., ${\sf T}_{\mu }={\sf T}%
^{\left( n_{\mu },N_{\mu }\right) }$, $n_{\mu },N_{\mu }\in {\Bbb N}$, a 
{\em mixing} of them by means of glueing the lattices on which they are
defined, in such a way that resulting mixed model can be solved by means of
algebraic Bethe ansatz technics.

\section{The glueing process}

For any pair $\left( R_{n},R_{m}\right) $ of matrices $\left( \ref{r}\right) 
$, there exist an associated quadratic algebra ${\rm A}\left(
R_{n}:R_{m}\right) $. They are called rectangular quantum matrix algebras 
\cite{mm}. There are also parameter dependent versions, the algebras ${\rm YB%
}_{n,m}$, generated by indeterminates ${\sf T}_{i}^{j}\left( x\right) $, $%
1\leq i\leq n$, $1\leq j\leq m$ and $x\in {\Bbb C}$, and defined by the
quadratic relations 
\begin{equation}
\left[ R_{n}\left( x/y\right) \right] _{ij}^{kl}\,\noindent \noindent {\sf T}%
_{k}^{r}\left( x\right) \,\noindent \noindent {\sf T}_{l}^{s}\left( y\right)
={\sf T}_{j}^{l}\,\noindent \noindent \left( y\right) \,{\sf T}%
_{i}^{k}\left( x\right) \,\noindent \noindent \left[ R_{m}\left( x/y\right) %
\right] _{kl}^{rs},  \label{rtt}
\end{equation}
$1\leq i,j\leq n$, $1\leq r,s\leq m$. Obviously, ${\rm YB}_{n,n}={\rm YB}%
_{n} $. In the same way as for the constant case \cite{mm}\cite{maj}, there
exist homomorphisms 
\begin{equation}
\Delta _{p}:{\rm YB}_{n,m}\rightarrow {\rm YB}_{n,p}\otimes {\rm YB}%
_{p,m};\;\;\;n,m,p\in {\Bbb N};  \label{coco}
\end{equation}
inherited from the cocomposition notion of the internal {\em coHom} objects,
enjoying the coassociativity property $\left( \Delta _{p}\otimes id\right)
\Delta _{r}=\left( id\otimes \Delta _{r}\right) \Delta _{p}$ \cite{hugo}. In
the $n=m=p$ cases, these reduce to the usual comultiplication maps [see Eq. $%
\left( \ref{cs}\right) $]. In particular, we have morphisms 
\[
{\rm YB}_{m}\rightarrow {\rm YB}_{m,n}\otimes {\rm YB}_{n}\otimes {\rm YB}%
_{n,m}\otimes {\rm YB}_{m} 
\]
for all $n,m$. Now, consider pure monodromy matrices ${\sf T}^{\left(
n,N\right) }$ and ${\sf T}^{\left( m,P\right) }$ related to ${\rm YB}_{n}$
and ${\rm YB}_{m}$, and representations $\lambda $ and $\beta $ of ${\rm YB}%
_{m,n}$ and ${\rm YB}_{n,m}$, respectively, where $\lambda $ and $\beta $
denote rectangular matrices whose coefficients are representative of the
corresponding generator algebra elements. Mentioned morphism implies $%
\lambda \stackrel{.}{\otimes }{\sf T}^{\left( n,N\right) }\stackrel{.}{%
\otimes }\beta \stackrel{.}{\otimes }{\sf T}^{\left( m,P\right) }$ gives a
representation of ${\rm YB}_{m}$. As we do not want to add new degrees of
freedom others than the related to the original models ${\sf T}^{\left(
n,N\right) }$ and ${\sf T}^{\left( m,P\right) }$, we ask $\lambda $ and $%
\beta $ to be constant (i.e., spectral parameter independent) 1-dimensional
representations. In this case $\lambda \stackrel{.}{\otimes }{\sf T}^{\left(
n,N\right) }\stackrel{.}{\otimes }\beta \stackrel{.}{\otimes }{\sf T}%
^{\left( m,P\right) }$ gives an operator on $\left( {\Bbb C}^{n}\right)
^{\otimes N}\otimes \left( {\Bbb C}^{m}\right) ^{\otimes P}$, which we shall
call the {\em glueing} of ${\sf T}^{\left( n,N\right) }$ and ${\sf T}%
^{\left( m,P\right) }$ through matrices $\lambda $ and $\beta $. It is worth
remarking that this is not the glueing operation defined in \cite{mm}.
Physically, $\lambda $ and $\beta $ define vertices with statistical weights 
\[
\lambda _{a}^{b}\;\;and\;\;\beta _{c}^{d},\;\;1\leq a,d\leq
m\;\;and\;\;1\leq b,c\leq n. 
\]
In general, for a family of pure monodromy matrices as described above, we
can define a {\em mixing} of them, namely 
\begin{equation}
{\sf T}^{mix}=\lambda _{0}\stackrel{.}{\otimes }{\sf T}_{0}\stackrel{.}{%
\otimes }\lambda _{1}\stackrel{.}{\otimes }{\sf T}_{1}\stackrel{.}{\otimes }%
...\stackrel{.}{\otimes }\lambda _{K-1}\stackrel{.}{\otimes }{\sf T}_{K-1},
\label{mvm}
\end{equation}
where each $\lambda _{\mu }$ is a constant 1-dimensional representation of
the rectangular YB algebra ${\rm YB}_{n_{\mu -1},n_{\mu }}$ (mod $K$).
Graphically, 
\[
\begin{array}{c}
\left[ {\sf T}_{a}^{b\;mix}\right] _{I_{0},...,I_{K-1}}^{J_{0},...,J_{K-1}}=%
{\small 
\begin{diagram}[size=1.5em,height=1.3em,shortfall=0em,midshaft]
    &                &             &                     &         {}               &                      &            &                           &          {}             &                         &             &                            &            {}                &                  & \\
    &                &             &                     &  \uLine_{J_{0}} &                      &            &                            & \uLine_{J_{1}} &                         &             &                            &  \uLine^{J_{K-1}}  &                 & \\
 \;& \rLine_{a}& \bullet & \rLine^{b_0}&    \bullet             & \rLine_{b_1} & \bullet & \rLine^{\;b_2\;} &       \bullet         &\rLine_{\;b_3\;}&\;...\;\; &\rLine_{b_{2K}} &       \bullet             & \rLine^{b}&{} \\
    &                &             &                     &   \uLine^{I_{0}} &                      &            &                            &\uLine^{I_{1}}   &                         &             &                            &  \uLine_{I_{K-1}}  &                  & \\
    &                &             &                     &          {}              &                      &            &                            &          {}             &                         &             &                            &            {}               &                   & \\
 \end{diagram}%
%
}\;{\small ;} \\ 
\\ 
\left[ {\sf T}_{\mu }\right] _{aI}^{bJ}={\small 
\begin{diagram}[size=1.5em,height=1.3em,shortfall=0em,midshaft]
 &              &                &                & \\
 &              & \uLine^J &                & \\
 & \rLine_a&   \bullet   & \rLine^b & \\
 &               & \uLine_I &                & \\
 &               &               &                & \\
 \end{diagram}%
%
}\;;{\small \;\;\;\;}\left[ \lambda _{\mu }\right] _{a}^{b}={\small 
\begin{diagram}[size=1.5em,shortfall=0em,midshaft]
& \rLine_a& \bullet & \rLine^b & \\
 \end{diagram}%
%
}\;,
\end{array}
\]
being $I_{\mu }$ and $J_{\mu }$ multi-indices for spaces $\left( {\Bbb C}%
^{n_{\mu }}\right) ^{\otimes N_{\mu }}$ on which each ${\sf T}_{\mu }$ acts.
Since the quadratic relations $\left( \ref{rtt}\right) $ and the
cocomposition maps $\left( \ref{coco}\right) $, one may see that ${\sf T}%
^{mix}$ provides a representation of ${\rm YB}_{n_{K-1}}$. This is a direct
consequence of the algebra map 
\begin{equation}
{\rm YB}_{n_{K-1}}\rightarrow {\rm YB}_{n_{K-1},n_{0}}\otimes {\rm YB}%
_{n_{0}}\otimes {\rm YB}_{n_{0},n_{1}}\otimes ...\otimes {\rm YB}%
_{n_{K-2},n_{K-1}}\otimes {\rm YB}_{n_{K-1}}.  \label{map}
\end{equation}
Of course, these representations are highly reducible in general, as we
shall see later.

\subsection{Constant one dimensional representations of ${\rm YB}_{n,m}$}

Representations $\lambda _{\mu }$ appearing in $\left( \ref{mvm}\right) $
match exactly with 1-dimensional representations of ${\rm A}\left(
R_{n}:R_{m}\right) $, i.e., rectangular matrices $\lambda \in Mat\left[
n\times m\right] $ in ${\Bbb C}$ such that 
\begin{equation}
\left[ R_{n}\right] _{ij}^{kl}\,\noindent \noindent \lambda
_{k}^{r}\,\noindent \noindent \lambda _{l}^{s}=\lambda _{j}^{l}\,\noindent
\noindent \,\lambda _{i}^{k}\,\noindent \noindent \left[ R_{m}\right]
_{kl}^{rs};\;\;1\leq i,j\leq n,\;1\leq r,s\leq m.  \label{nm}
\end{equation}
We are considering the same parameter $q\neq 0,1$ for all involved $R$%
-matrices. Otherwise, the only solution to $\left( \ref{nm}\right) $ is the
trivial one. Using explicit form of $R_{n}$ given in $\left( \ref{r}\right) $%
, last equation is equivalent to 
\[
\begin{array}{ll}
\lambda _{i}^{r}\,\lambda _{j}^{r}=0, & \;\;1\leq r\leq m,\;1\leq i\leq
j\leq n, \\ 
\lambda _{i}^{r}\,\lambda _{i}^{s}=0, & \;\;1\leq r<s\leq m,\;1\leq i\leq n,
\\ 
\lambda _{i}^{r}\,\lambda _{j}^{s}=0, & \;\;1\leq r<s\leq m,\;1\leq j<i\leq
n.
\end{array}
\]
First and second lines imply coefficients of $\lambda $ in a given column
and row, respectively, are null except for almost one of them. Last line
says, if $\lambda _{i}^{j}\neq 0$ then all coefficients $\lambda _{a}^{b}$
with $i<a$, $b<j$, and with $a<i$, $j<b$, are null. Thus, each solution $%
\lambda $ of $\left( \ref{nm}\right) $ is a diagonal matrix to which columns
and rows of zeros were added. From that it follows immediately the set of
solutions for all $m,n$ form a semigroupoid, or a category, generated by the
abelian groups ${\cal D}_{n}$ of invertible $n\times n$ diagonal matrices,
and also by matrices $\sigma _{i}^{n}\in Mat\left[ n\times \left( n+1\right) 
\right] $ and $\partial _{i}^{n}\in Mat\left[ \left( n+1\right) \times n%
\right] $, $i=1,...,n+1$, $n\in {\Bbb N}$, given by 
\[
\sigma _{i}^{n}=\left[ 
\begin{array}{cc}
Id_{\left( i-1\right) \times \left( i-1\right) } & O_{\left( i-1\right)
\times \left( n-i+2\right) } \\ 
O_{1\times \left( i-1\right) } & O_{1\times \left( n-i+2\right) } \\ 
O_{\left( n-i\right) \times \left( i-1\right) } & Id_{\left( n-i\right)
\times \left( n-i+2\right) }
\end{array}
\right] ,\;\partial _{i}^{n}=\left[ 
\begin{array}{ccc}
Id_{\left( i-1\right) \times \left( i-1\right) } & O_{\left( i-1\right)
\times 1} & O_{\left( i-1\right) \times \left( n-i\right) } \\ 
O_{\left( n-i+2\right) \times \left( i-1\right) } & O_{\left( n-i+2\right)
\times 1} & Id_{\left( n-i+2\right) \times \left( n-i\right) }
\end{array}
\right] , 
\]
being $O_{n,m}$ the $n\times m$ null matrix. In fact, a general solution of
Eq. $\left( \ref{nm}\right) $ has the form 
\begin{equation}
\begin{array}{c}
\lambda =\partial _{j_{b}}^{n-1}\,...\,\,\partial _{j_{1}}^{k}\,D\,\sigma
_{i_{1}}^{k}\,...\,\,\sigma _{i_{a}}^{m-1}\in Mat\left[ n\times m\right] ;
\\ 
a,b\geq 0,\;\;m-a=n-b=k\geq 0,
\end{array}
\label{el}
\end{equation}
with $i_{1}\leq ...\leq i_{a}\leq m$, $j_{1}\leq ...\leq j_{b}\leq n$, and $%
D\in {\cal D}_{k}$. If $a$ (resp. $b$) is equal to zero, then factors of
type $\sigma $ (resp. $\partial $) do not appear. Such a solution has $k$
non null entries equal to the diagonal elements of $D$, a number $a$ of null
columns in positions $i_{1},...,i_{a}$, and $b$ null rows in positions $%
j_{1},...,j_{b}$. Note that $%
\mathop{\rm rank}%
\lambda =k$.

Matrices $\sigma _{i}^{n}$ and $\partial _{i}^{m}$, which give solutions to $%
\left( \ref{nm}\right) $ for $m=n+1$ and $n=m+1$, respectively, are related
each other by matrix transposition, i.e., $\partial _{i}^{n}=\left( \sigma
_{i}^{n}\right) ^{t}$, and enjoy relations 
\[
\begin{array}{ll}
\sigma _{j}^{n-1}\,\sigma _{i}^{n}=\sigma _{i\,}^{n-1}\sigma _{j+1}^{n}, & 
\;\;\;i\leq j; \\ 
\partial _{i}^{n+1}\,\partial _{j}^{n}=\partial _{j+1}^{n+1}\,\partial
_{i}^{n}, & \;\;\;i\leq j; \\ 
\sigma _{j}^{n+1}\,\partial _{i}^{n+1}=\partial _{i\,}^{n}\,\sigma
_{j-1}^{n}, & \;\;\;i<j; \\ 
\sigma _{j}^{n}\,\partial _{i}^{n}=Id, & \;\;\;i=j; \\ 
\sigma _{j}^{n+1}\,\partial _{i}^{n+1}=\partial _{i-1}^{n}\,\sigma _{j}^{n},
& \;\;\;i\geq j+1.
\end{array}
\]
In spite of these relations, they do {\bf not} define the simplicial
category. Note that, for instance, $\sigma _{j}^{n}\,\partial _{j+1}^{n}\neq
Id$. Nevertheless we name $\Delta =\bigvee_{n,m\in {\Bbb N}}\Delta _{n,m}$
the category formed out by them.

On the other hand, as it is well known, the group of diagonal matrices $%
{\cal D}_{n}$ defines precisely the symmetry group of $R_{n}$, given by
matrices $D\in GL\left( n\right) $ such that $\left[ R_{n},D\otimes D\right]
=0$. Moreover, they are also the symmetry group of $R_{n}\left( x\right) $
or $R_{n}\left( x,y\right) =R_{n}\left( x/y\right) $. Let us mention that,
when $R_{n}\left( x,y\right) $ is changed by a similarity transformation $%
Q\left( x\right) \otimes Q\left( y\right) $ such that $Q_{k}^{l}\left(
x\right) =\delta _{k}^{l}\,x^{2l/n}$, the group enlarges to ${\cal D}%
_{n}\times {\Bbb Z}_{n}$ \cite{dev}. This is why systems related to such $R$%
-matrices were called ${\Bbb Z}_{n}$-symmetric vertex models \cite{zn}.

Elements $D\in {\cal D}_{n}$ give rise to multiparametric solutions $\left(
id\otimes D\right) ^{-1}\,R_{n}\,\left( D\otimes id\right) $ of the YB
equation \cite{man}, and related twist transformations of original quantum
groups \cite{maj}\cite{drin}. Associated integrable models, which differ
from the original ones by a twisting of the boundary conditions, were
described in \cite{foer}. We shall see latter that also in mixed models the
role of matrices $D$ is to make a twist on the boundary conditions.

The commutation relations between elements of ${\cal D}=\bigvee_{n\in {\Bbb N%
}}{\cal D}_{n}$ and $\Delta $ can be written 
\begin{equation}
\begin{array}{c}
D\,\sigma _{i}^{n}=\sigma _{i}^{n}\,D_{i}^{+},\;\;\;\;\partial
_{i}^{n}\,D=D_{i}^{+}\,\partial _{i}^{n}, \\ 
\sigma _{i}^{n-1}\,D=D_{i}^{-}\,\sigma _{i}^{n-1},\;\;\;\;D\,\partial
_{i}^{n-1}=\partial _{i}^{n-1}\,D_{i}^{-},
\end{array}
\label{cr}
\end{equation}
being $D_{i}^{+}=diag\left( d_{1},...,d_{i-1},1,d_{i},...,d_{n}\right) $ and 
$D_{i}^{-}=diag\left( d_{1},...,d_{i-1},d_{i+1},...,d_{n}\right) $ whenever $%
D=diag\left( d_{1},...,d_{n}\right) \in {\cal D}_{n}$. It is worth
mentioning that $\lambda $ can also be expressed as a product $\lambda
=\zeta \,D^{\prime }$, where $\zeta \in \Delta $ is obtained from $\lambda $
by taking $D=Id$, and $D^{\prime }\in {\cal D}_{n}$ is the result of passing 
$D$ to the right, through matrices $\,\sigma $'s, using commutation rules $%
\left( \ref{cr}\right) $.

\subsection{Equivalent forms for a mixed monodromy matrix}

From results of last section, it is clear that any mixed model has a
monodromy matrix 
\begin{equation}
\lambda _{0}\cdot {\sf R}_{0}\stackrel{.}{\otimes }\lambda _{1}\cdot {\sf R}%
_{1}\stackrel{.}{\otimes }...\stackrel{.}{\otimes }\lambda _{N-1}\cdot {\sf R%
}_{N-1},  \label{ca}
\end{equation}
with ${\sf R}_{\nu }={\sf R}^{n_{\nu }}$ for some related dimension $n_{\nu
} $, and $\lambda _{\nu }=\zeta _{\nu }\,D_{\nu }$, where $\zeta _{\nu }$ is
in $\Delta _{n_{\nu -1},n_{\nu }}$ and $D_{\nu }$ in ${\cal D}_{n_{\nu }}$.
Here $\nu \in {\Bbb Z}_{N}$. Equation $\left( \ref{mvm}\right) $ corresponds
to the case in which there exist $K$ numbers $N_{\mu }$, $\mu \in {\Bbb Z}%
_{K}$, giving a partition of $N$ and such that 
\begin{equation}
{\sf R}_{M_{\mu }}=...={\sf R}_{M_{\mu }+N_{\mu }-1}={\sf R}^{n_{M_{\mu
}}}\;\;and\;\;\;\lambda _{\nu }=Id_{n_{\nu }}\;for\;\nu \neq
M_{0},...,M_{K-1},  \label{lt}
\end{equation}
being $M_{0}=0$ and $M_{\mu }=\sum_{\sigma =0}^{\mu -1}N_{\sigma }$ for $%
1\leq \mu \leq K-1$. Furthermore, Eq. $\left( \ref{ca}\right) $ can be
brought to an equivalent form 
\begin{equation}
{\sf T}^{mix}=\zeta _{0}\cdot {\sf R}_{0}\stackrel{.}{\otimes }\zeta
_{1}\cdot {\sf R}_{1}\stackrel{.}{\otimes }...\stackrel{.}{\otimes }\zeta
_{N-1}\cdot {\sf R}_{N-1}\,\Upsilon ,  \label{can}
\end{equation}
where $\Upsilon $ is an element of ${\cal D}_{n_{N-1}}$. To see that, let us
define matrices $D_{\nu }^{\left( k\right) }\in {\cal D}_{n_{k}}$, $k\in 
{\Bbb Z}_{N}$, by 
\[
D_{\nu }^{\left( k\right) }=\left\{ 
\begin{array}{ll}
Id_{n_{k}}, & 0\leq k<\nu , \\ 
D_{\nu }, & k=\nu ,
\end{array}
\right. 
\]
and for each $k>\nu $ by the solution of $D_{\nu }\,\zeta _{\nu +1}\,\zeta
_{\nu +2}\,...\,\zeta _{k}=\zeta _{\nu +1}\,\zeta _{\nu +2}\,...\,\zeta
_{k}\,D_{\nu }^{\left( k\right) }$. We mean by `the solution' of last
equation the inversible diagonal matrix $D_{\nu }^{\left( k\right) }$ that
arises when passing, in the first member, the matrix $D_{\nu }$ to the right
using $\left( \ref{cr}\right) $. Then $\left( \ref{ca}\right) $ and $\left( 
\ref{can}\right) $ are similar through $\Pi _{\nu \in {\Bbb Z}_{N}}\otimes
_{k\in {\Bbb Z}_{N}}D_{\nu }^{\left( k\right) }$, and $\Upsilon =\Pi _{\nu
\in {\Bbb Z}_{N}}D_{\nu }^{\left( N-1\right) }$.

In other words, every mixed model is physically equivalent to a twisted
version of another one whose corresponding matrices $\lambda _{\nu }$ are in
the category $\Delta $. The role of matrices $D$'s is to implement a
twisting of the boundary conditions. Accordingly, we can describe each
mixing ${\sf T}^{mix}$ in terms of a family of elements $\zeta _{\nu }\in
\Delta _{n_{\nu -1},n_{\nu }}$, the {\em glueing matrices}, and a diagonal
matrix $\Upsilon =diag\left( \tau _{1},...,\tau _{n_{N-1}}\right) $ of $%
{\cal D}_{n_{N-1}}$, the {\em boundary matrix}.

Another useful expression for the monodromy matrices of these models can be
given from the following observation. Any matrix $\lambda \in \Delta _{n,m}$
of rank $k$ (note that $k\leq m,n$) may be written $\lambda =P\,\widehat{%
\lambda }\,P^{\prime }$, where $\widehat{\lambda }=\partial
_{n}^{n-1}\,...\,\,\partial _{k+1}^{k}\,\sigma _{k+1}^{k}\,...\,\,\sigma
_{m}^{m-1}\in Mat\left[ n\times m\right] $, that is, 
\begin{equation}
\widehat{\lambda }=\left( 
\begin{array}{cc}
Id_{k} & O_{k\times \left( m-k\right) } \\ 
O_{\left( n-k\right) \times k} & O_{\left( n-k\right) \times \left(
m-k\right) }
\end{array}
\right) ,  \label{hat}
\end{equation}
and $P,P^{\prime }$ are appropriate permutations. More precisely, if 
\[
\lambda =\partial _{j_{b}}^{n-1}\,...\,\,\partial _{j_{1}}^{k}\,\sigma
_{i_{1}}^{k}\,...\,\sigma _{i_{a}}^{m-1},\;i_{1}\leq ...\leq i_{a}\leq
m,\;j_{1}\leq ...\leq j_{b}\leq n 
\]
[see Eq. $\left( \ref{el}\right) $], then we can choose, for instance, $P\in
Mat\left[ n\right] $ and $P^{\prime }\in Mat\left[ m\right] $ to be 
\[
P=C_{j_{b},n}\,C_{j_{b-1},n}\,...\,C_{j_{1},n}\;\;\;and\;\;P^{\prime
}=C_{i_{1},m}\,...\,C_{i_{a-1},m}\,C_{i_{a},m}, 
\]
respectively, being $C_{r,s}$, $r\leq s$, the matrix that acting on the
right (resp. left) makes a cyclic permutation sending $s$-th column (resp.
row) to $r$-th one, and acts as an identity for the rest of columns (resp.
rows). Hence, for a given family of glueing matrices we have $\zeta _{\nu
}=P_{\nu }\,\widehat{\zeta }_{\nu }\,P_{\nu }^{\prime }$. Introducing last
expression for $\zeta _{\nu }$ into Equation $\left( \ref{can}\right) $, and
making a similarity transformation $\otimes _{\nu \in {\Bbb Z}_{N}}P_{\nu
}^{\prime }$, an equivalent system 
\begin{equation}
\begin{array}{l}
\widetilde{{\sf T}}^{mix}=P_{0}\,\,\widehat{\zeta }_{0}\cdot \widetilde{{\sf %
R}}_{0}\cdot Q_{0}\,\stackrel{.}{\otimes }\widehat{\zeta }_{1}\cdot 
\widetilde{{\sf R}}_{1}\cdot Q_{1}\stackrel{.}{\otimes }... \\ 
\;\;\;\;\;\;\;\;\;\;\;\;...\stackrel{.}{\otimes }\widehat{\zeta }_{N-2}\cdot 
\widetilde{{\sf R}}_{N-2}\cdot Q_{N-2}\stackrel{.}{\otimes }\widehat{\zeta }%
_{N-1}\cdot \widetilde{{\sf R}}_{N-1}\,P_{N-1}^{\prime }\,\Upsilon ,
\end{array}
\label{cang}
\end{equation}
where $Q_{\nu }=P_{\nu }^{\prime }\,P_{\nu +1}$ and $\widetilde{{\sf R}}%
_{\nu }=\left( P_{\nu }^{\prime }\otimes P_{\nu }^{\prime }\right) \,{\sf R}%
_{\nu }\,\left( P_{\nu }^{\prime }\otimes P_{\nu }^{\prime }\right) ^{-1}$,
follows. That is, 
\begin{equation}
\widetilde{{\sf T}}^{mix}=\left( \otimes _{\nu \in {\Bbb Z}_{N}}P_{\nu
}^{\prime }\right) \,{\sf T}^{mix}\,\left( \otimes _{\nu \in {\Bbb Z}%
_{N}}P_{\nu }^{\prime }\right) ^{-1}.  \label{tr}
\end{equation}
It can be seen for an arbitrary permutation that 
\begin{equation}
\widetilde{{\sf R}}_{a}^{a}\left( x\right) ={\sf R}_{a}^{a}\left( x\right)
,\;\;and\;\;\widetilde{{\sf R}}_{a}^{b}\left( x\right) =x^{2\varepsilon
_{ab}}\,{\sf R}_{a}^{b}\left( x\right) \;\;\;for\;\;a\neq b,  \label{re}
\end{equation}
where coefficients $\varepsilon _{ab}$ takes values $-1,0,1$ depending on
the considered permutation. We shall show in the next section that mixed
models with 
\begin{equation}
P_{0}=P_{N-1}^{\prime }=Id_{n_{N-1}},\;\;and\;\;Q_{\nu }=Id_{n_{\nu
}},\forall \nu \in {\Bbb Z}_{N-1},  \label{co}
\end{equation}
are solvable by means of algebraic Bethe ansatz technics [actually, $\left( 
\ref{co}\right) $ can be slightly relaxed and ask $Q_{\nu }=Id_{n_{\nu
}},\forall \nu \in {\Bbb Z}_{N}$, instead]. Furthermore, we shall see
complete integrability implies transfer matrix obtained from $\left( \ref
{cang}\right) $ is similar to the trace of 
\begin{equation}
{\sf T}^{mix}=\widehat{\zeta }_{0}\cdot {\sf R}_{0}\stackrel{.}{\otimes }%
\widehat{\zeta }_{1}\cdot {\sf R}_{1}\stackrel{.}{\otimes }...\stackrel{.}{%
\otimes }\widehat{\zeta }_{N-1}\cdot {\sf R}_{N-1}\,\Upsilon .  \label{canm}
\end{equation}
Thus, we can solve all vertex models with glueing matrices $\zeta _{\nu
}=P_{\nu }\,\widehat{\zeta }_{\nu }\,P_{\nu }^{\prime }$ satisfying $\left( 
\ref{co}\right) $ by solving those with ${\sf T}^{mix}$ given in Equation $%
\left( \ref{canm}\right) $. In addition, all $\widehat{\zeta }_{\nu }$ can
be supposed to have the same rank.

\section{Integrability of mixed vertex models}

In order to show exact solvability of these models (or unless of a subclass
of them), since needed commutation rules follow from map $\left( \ref{map}%
\right) $, we must prove there exists a suitable set of pseudovacuum vectors
for ${\sf T}^{mix}$ from which all its eigenstates and corresponding
eigenvalues can be constructed. In other terms, using block form 
\[
{\sf T}^{mix}=\left[ 
\begin{array}{cc}
{\sf A}^{mix} & {\sf B}_{j}^{mix} \\ 
{\sf C}_{i}^{mix} & {\sf D}_{ij}^{mix}
\end{array}
\right] ;\;\;\;1\leq i,j\leq n_{N-1}-1, 
\]
we look for elements $\Phi \in {\cal H}^{mix}=\otimes _{\nu \in {\Bbb Z}_{N}}%
{\Bbb C}^{n_{\nu }}$ which are eigenvectors of ${\sf A}^{mix}$ and of each
diagonal entry ${\sf D}_{ii}^{mix}$, such that ${\sf C}_{i}^{mix}\,\Phi =0$,
and ${\sf D}_{ij}^{mix}\,\Phi =0$ for $i\neq j$. In this way, we build up
recursively all eigenvalues and eigenstates by applying repeatedly operators 
${\sf B}_{j}^{mix}$ to the mentioned vectors. Completeness problem will be
studied separately. Of course, the smaller the rank of involved glueing
matrices, the smaller the set of monomials in ${\sf B}_{j}^{mix}$ and the
bigger the number of pseudovacuum vectors we need to construct the complete
set of eigenstates. In the singular case for which $%
\mathop{\rm rank}%
\zeta _{\nu }=0$ for some $\nu $, we have ${\sf T}^{mix}=0$ and accordingly
every vector of ${\cal H}^{mix}$ is trivially a pseudovacuum vector, and no
creation operator is needed in order to diagonalize the transfer matrix $%
{\sf t}^{mix}$. Note in this case, operators ${\sf B}_{j}^{mix}$ are null.
Thus we can have pseudovacuum vectors which are annihilated by operators $%
{\sf B}_{j}^{mix}$ and still be able to build up an eigenstate basis for $%
{\sf t}^{mix}$.

We actually show exact solvability for a particular class of mixed models.
Concretely, we concentrate ourself in monodromy matrices whose related
glueings $\zeta _{\nu }$ satisfy Eq. $\left( \ref{co}\right) $.

\subsection{The pseudovacuum subspace}

Let us first consider the mixed models with monodromy matrices ${\sf T}%
^{mix} $ defined by Equation $\left( \ref{canm}\right) $, that is, each $%
\zeta _{\nu }=\widehat{\zeta }_{\nu }$ is of the form $\left( \ref{hat}%
\right) $. They are a particular case of those with glueing matrices
satisfying $\left( \ref{co}\right) $. At the end of this section the general
case will be addressed. In order to simplify our calculations, we shall
suppose 
\begin{equation}
m\doteq 
\mathop{\rm rank}%
\zeta _{0}\leq 
\mathop{\rm rank}%
\zeta _{\nu }\;\;\;for\;all\;\;\nu \in {\Bbb Z}_{N}.  \label{coa}
\end{equation}
This can be reached by a similarity transformation cyclically permuting
tensor factors of the linear space ${\cal H}^{mix}$. Of course, $m\leq 
\mathop{\rm rank}%
\zeta _{\nu }\leq min\left\{ n_{\nu -1},n_{\nu }\right\} $ (mod\ $N$). Also,
we suppose $m>0$, since for $m=0$ diagonalization of ${\sf T}^{mix}$ is
immediate. Note that $\left( \ref{coa}\right) $ implies 
\begin{equation}
{\sf T}_{a}^{b\;mix}=0\;\;\;for\;\;a>m,  \label{I}
\end{equation}
and in particular 
\begin{equation}
\begin{array}{l}
{\sf t}^{mix}=\sum_{a=1}^{n_{N-1}}{\sf T}_{a}^{a\;mix}=\sum_{a=1}^{m}{\sf T}%
_{a}^{a\;mix}={\sf A}^{mix}+\sum_{i=1}^{m-1}{\sf D}_{ii}^{mix}.
\end{array}
\label{tm}
\end{equation}
For $a\leq m$ we have (non sum over $b$) 
\begin{equation}
{\sf T}_{a}^{b\;mix}=\tau _{b}\,\left[ {\sf R}^{n_{0}}\right]
_{a}^{c_{1}}\otimes \left[ {\sf R}^{n_{1}}\right] _{c_{1}}^{c_{2}}\,...\,%
\left[ {\sf R}^{n_{N-2}}\right] _{c_{N-2}}^{c_{N-1}}\otimes \left[ {\sf R}%
^{n_{N-1}}\right] _{c_{N-1}}^{b},  \label{T}
\end{equation}
where sum over each $c_{\nu }$ is in the interval $1\leq c_{\nu }\leq 
\mathop{\rm rank}%
\zeta _{\nu }$.

Let us indicate by $e_{1},...,e_{n}$ the elements of the canonical basis of $%
{\Bbb C}^{n}$. Then ${\cal H}^{mix}$ is spanned by vectors of the form $%
e_{f_{0}}\otimes ...\otimes e_{f_{N-1}}$, which can be identified with an
obvious subset $\digamma $ of functions $f:{\Bbb Z}_{N}\rightarrow {\Bbb N}%
:\nu \mapsto f_{\nu }$. In particular, given $f\in \digamma $, we denote $%
\Omega ^{f}$ the corresponding vector of ${\cal H}^{mix}$. We shall show
there exists a set of pseudovacuum vectors, on which algebraic Bethe ansatz
will be applied, labeled by the subset $\digamma _{0}$ of functions 
\begin{equation}
f\in \digamma \;\;/\;\;%
\mathop{\rm Image}%
f\subset \left\{ 1\right\} \cup \left\{ n\in {\Bbb N}:n>m\right\} .
\label{set}
\end{equation}
More precisely, there exist vectors $\Phi ^{f}\in {\cal H}^{mix}$, $f\in
\digamma _{0}$, expanding a space 
\begin{equation}
{\cal H}_{0}\doteq span\left\{ \Omega ^{f}\in {\cal H}^{mix}:f\in \digamma
_{0}\right\} \subset {\cal H}^{mix},  \label{hcero}
\end{equation}
namely the pseudovacuum subspace, and fulfilling 
\begin{equation}
\begin{array}{ccc}
& {\sf A}^{mix}\,\Phi ^{f}=\tau _{1}\,d\,\prod_{\nu \in f^{-1}\left(
1\right) }G\left( x/\alpha _{\nu }\right) \,\Phi ^{f}, &  \\ 
& {\sf D}_{ii}^{mix}\,\Phi ^{f}=\tau _{i+1}\,d\,\Phi ^{f}\;\;\left(
i<m\right) ,\;\;\;\;{\sf D}_{i\neq j}^{mix}\,\Phi ^{f}={\sf C}%
_{i}^{mix}\,\Phi ^{f}=0, & 
\end{array}
\label{II}
\end{equation}
being $d=%
\mathop{\textstyle\prod}%
_{\nu \in {\Bbb Z}_{N}}1/G\left( x/\alpha _{\nu }\right) $ and $G\left(
x\right) =\left( xq-1/qx\right) /\left( x-1/x\right) $. In particular for
the {\em equal rank case}, i.e., if $%
\mathop{\rm rank}%
\zeta _{\nu }=m$ for all $\nu $, then $\Phi ^{f}=\Omega ^{f}$. Note that $%
{\cal H}_{0}\subset \ker \noindent \noindent {\sf C}_{i}^{mix}$ $\forall i$.

We also show 
\begin{equation}
{\sf B}_{j}^{mix}\,\Phi ^{f}\neq c\,\Phi ^{f}\;\;\;if\;\;1\in 
\mathop{\rm Image}%
f;\;\;otherwise,\;\;{\sf B}_{j}^{mix}\,\Phi ^{f}=0;  \label{III}
\end{equation}
i.e., each ${\sf B}_{j}^{mix}$ creates new states when $f^{-1}\left(
1\right) \neq \emptyset $. Using that we construct a set of Bethe vectors
from each $\Phi ^{f}$, with $j$ from $1$ to $m$ and $f^{-1}\left( 1\right)
\neq \emptyset $, and generate in this way all eigenstates of the transfer
matrix.

To find the vectors $\Phi ^{f}$ we need some previous results.

\subsubsection{The action of ${\sf T}^{mix}$ on vectors $\Omega ^{f}$}

Let us evaluate the entries ${\sf T}_{a}^{b\;mix}$ on each vector $\Omega
^{f}$. From Eq. $\left( \ref{I}\right) $ it follows that ${\sf T}%
_{a}^{b\;mix}\,\Omega ^{f}=0$ for all $a>m$. So we only consider $a\leq m$.
As usual \cite{dev}, we normalize operators ${\sf R}_{a}^{b}={\sf R}^{n_{\nu
}}\,_{a}^{b}\left( x/\alpha _{\nu }\right) :{\Bbb C}^{n_{\nu }}\rightarrow 
{\Bbb C}^{n_{\nu }}$ in such a way that on the canonical basis of ${\Bbb C}%
^{n_{\nu }}$ 
\begin{equation}
{\sf R}_{a}^{b}\,e_{k}=\left\{ 
\begin{array}{ll}
\delta _{a}^{b}\,/G\left( x/\alpha _{\nu }\right) \,e_{k}, & k\neq a, \\ 
\left( \delta _{a}^{b}+\left( 1-\delta _{a}^{b}\right) \,c_{%
\mathop{\rm sg}%
\left( b-a\right) }\left( x/\alpha _{\nu }\right) \right) \,e_{b}, & k=a,
\end{array}
\right.  \label{compa}
\end{equation}
being $c_{\pm }\left( x\right) =\left( q-1/q\right) \,\,x^{\pm 1}/\left(
xq-1/qx\right) $. Then, from $\left( \ref{T}\right) $ and the first part of $%
\left( \ref{compa}\right) $, it follows that 
\begin{equation}
\begin{array}{l}
{\sf T}_{a}^{b\;mix}\,\Omega ^{f}=\tau _{b}\,\delta _{a}^{b}\,%
\mathop{\textstyle\prod}%
_{\nu \in {\Bbb Z}_{N}}1/G\left( x/\alpha _{\nu }\right) \,\Omega ^{f}=\tau
_{a}\,\delta _{a}^{b}\,d\,\Omega ^{f}
\end{array}
\label{adb}
\end{equation}
if $a\notin 
\mathop{\rm Image}%
f$, i.e., if $f\left( \nu \right) \neq a$ for all $\nu \in {\Bbb Z}_{N}$. In
particular 
\begin{equation}
{\sf T}_{a}^{b\;mix}\,\Omega ^{f}=0\;\;\;if\;\;\;a\notin 
\mathop{\rm Image}%
f\cup \left\{ b\right\} .  \label{adbb}
\end{equation}
Also, if $f\in \digamma _{0}$ and $1<a\leq m$, since in this case $a\notin 
\mathop{\rm Image}%
f$ [see $\left( \ref{set}\right) $], we have that 
\begin{equation}
\begin{array}{l}
{\sf D}_{ii}^{mix}\,\Omega ^{f}={\sf T}_{i+1}^{i+1\;mix}\,\Omega ^{f}=\tau
_{i+1}\,d\,\Omega ^{f},\;\;for\;\;1\leq i<m, \\ 
{\sf D}_{ij}^{mix}\,\Omega ^{f}={\sf T}_{i+1}^{j+1\;mix}\,\Omega
^{f}=0,\;\;for\;\;1\leq i,j<m,\;i\neq j, \\ 
{\sf C}_{i}^{mix}\,\Omega ^{f}={\sf T}_{i+1}^{1\;mix}\,\Omega
^{f}=0,\;\;for\;\;1\leq i<m,
\end{array}
\label{eqI}
\end{equation}
putting $i=a-1$ in $\left( \ref{adb}\right) $. Otherwise, let $\sigma _{a}$
be the first integer such that $f\left( \sigma _{a}\right) =a$, that is, $%
f\left( \nu \right) \neq a$ for all $\nu <\sigma _{a}$ and $f\left( \sigma
_{a}\right) =a$. Let us write 
\begin{equation}
\Omega ^{f}=\Omega ^{g^{a}}\otimes e_{a}\otimes \Omega
^{f^{a}},\;\;\;with\;\;\;g^{a},f^{a}:{\Bbb Z}_{\sigma _{a}},{\Bbb Z}%
_{N-\sigma _{a}-1}\rightarrow {\Bbb N}.  \label{deco}
\end{equation}
If $\sigma _{a}=N-1$, we take $\Omega ^{f^{a}}$ equal to $1$. Then, using $%
\left( \ref{T}\right) $ and $\left( \ref{compa}\right) $ again (note that $a$
does not belong to $%
\mathop{\rm Image}%
g^{a}$) we have 
\begin{equation}
\begin{array}{c}
{\sf T}_{a}^{b\;mix}\,\Omega ^{f}=\sum_{i=1}^{%
\mathop{\rm rank}%
\zeta _{\sigma _{a}}}C_{a,i}\,\Omega ^{g^{a}}\otimes e_{i}\otimes \widehat{%
{\sf T}}_{i}^{b\;mix}\,\Omega ^{f^{a}}, \\ 
C_{a,i}=D_{a,i}%
\mathop{\textstyle\prod}%
_{\nu <\sigma _{a}}\left. 1\right/ G\left( x/\alpha _{\nu }\right)
;\;\;\;\;\;D_{a,i}=\delta _{a}^{i}+\left( 1-\delta _{a}^{i}\right) \,c_{%
\mathop{\rm sg}%
\left( i-a\right) }\left( \left. x\right/ \alpha _{\sigma _{a}}\right) .
\end{array}
\label{to}
\end{equation}
Here, operators $\widehat{{\sf T}}_{i}^{b\;mix}$ are given by the last $%
N-\sigma _{a}-1$ factors of ${\sf T}^{mix}$. From Eq. $\left( \ref{adbb}%
\right) $ applied to $\widehat{{\sf T}}_{i}^{b\;mix}\,\Omega ^{f^{a}}$,
since $C_{a,i}\neq 0$ for all $1\leq i\leq 
\mathop{\rm rank}%
\zeta _{\sigma _{a}}$, the non zero terms of $\left( \ref{to}\right) $ are
those with $i$ inside $I_{a}=\left\{ i\in 
\mathop{\rm Image}%
f^{a}\cup \left\{ b\right\} :i\leq 
\mathop{\rm rank}%
\zeta _{\sigma _{a}}\right\} $. If $b$ is the unique element of $I_{a}$ and $%
b\notin 
\mathop{\rm Image}%
f^{a}$, then 
\begin{equation}
\begin{array}{r}
{\sf T}_{a}^{b\;mix}\,\Omega ^{f}=\tau _{b}\,C_{a,b}\,\left( 
\mathop{\textstyle\prod}%
_{\nu \neq \sigma _{a}}\left. 1\right/ G\left( x/\alpha _{\nu }\right)
\right) \,\Omega ^{g^{a}}\otimes e_{b}\otimes \Omega ^{f^{a}} \\ 
=\tau _{b}\,C_{a,b}\,d\,G\left( x/\alpha _{\sigma _{a}}\right) \,\Omega
^{g^{a}}\otimes e_{b}\otimes \Omega ^{f^{a}}.
\end{array}
\label{s}
\end{equation}
Otherwise, suppose there exists $c_{1}\in I_{a}\cap 
\mathop{\rm Image}%
f^{a}$, and let $\sigma _{ac_{1}}$ be the first integer such that $%
f^{a}\left( \sigma _{ac_{1}}-\sigma _{a}\right) =c_{1}$. Then $\widehat{{\sf %
T}}_{c_{1}}^{b\;mix}\,\Omega ^{f^{a}}=\sum_{i\in
I_{ac_{1}}}C_{ac_{1},i}\,\Omega ^{g^{ac_{1}}}\otimes e_{i}\otimes \widehat{%
{\sf T}}_{i}^{b\;mix}\,\Omega ^{f^{ac_{1}}}$ with 
\[
\begin{array}{ccc}
& C_{ac_{1},i}=D_{c_{1},i}\,%
\mathop{\textstyle\prod}%
_{\sigma _{a}<\nu <\sigma _{ac_{1}}}1/G\left( x/\alpha _{\nu }\right)
,\;\;\;\Omega ^{f^{a}}=\Omega ^{g^{ac_{1}}}\otimes e_{c_{1}}\otimes \Omega
^{f^{ac_{1}}}, &  \\ 
& I_{ac_{1}}=\left\{ i\in 
\mathop{\rm Image}%
f^{ac_{1}}\cup \left\{ b\right\} :i\leq 
\mathop{\rm rank}%
\zeta _{\sigma _{ac_{1}}}\right\} . & 
\end{array}
\]
A recursive process easily follows and the generic term reads 
\[
\begin{array}{c}
\widehat{{\sf T}}_{c_{k}}^{b\;mix}\,\Omega
^{f^{ac_{1}...c_{k-1}}}=\sum_{i\in
I_{ac_{1}...c_{k}}}\,C_{ac_{1}...c_{k},i}\,\Omega
^{g^{ac_{1}...c_{k}}}\otimes e_{i}\otimes \widehat{{\sf T}}%
_{i}^{b\;mix}\,\Omega ^{f^{ac_{1}...c_{k}}},
\end{array}
\]
with $C_{ac_{1}...c_{k},i}=D_{c_{k},i}\,%
\mathop{\textstyle\prod}%
_{\sigma _{ac_{1}...c_{k-1}}<\nu <\sigma _{ac_{1}...c_{k}}}1/G\left(
x/\alpha _{\nu }\right) $. Of course, $c_{1}\in I_{a}\cap 
\mathop{\rm Image}%
f^{a}$, 
\begin{equation}
I_{ac_{1}c_{2}...c_{j-1}}=\left\{ i\in 
\mathop{\rm Image}%
f^{ac_{1}c_{2}...c_{j-1}}\cup \left\{ b\right\} :i\leq 
\mathop{\rm rank}%
\zeta _{\sigma _{ac_{1}...c_{j-1}}}\right\}  \label{ii}
\end{equation}
and $c_{j}\in I_{ac_{1}c_{2}...c_{j-1}}\cap 
\mathop{\rm Image}%
f^{ac_{1}c_{2}...c_{j-1}}$ for $2\leq j\leq k$. The process ends when $b\in
I_{ac_{1}...c_{k}}$, $b\notin 
\mathop{\rm Image}%
f^{ac_{1}...c_{k}}$, and we choose $c_{k+1}=b$. In this case 
\[
\begin{array}{l}
\widehat{{\sf T}}_{c_{k}}^{b\;mix}\,\Omega ^{f^{ac_{1}...c_{k-1}}}=\tau
_{b}\,C_{ac_{1}...c_{k},b}\,%
\mathop{\textstyle\prod}%
_{\nu >\sigma _{ac_{1}...c_{k}}}1/G\left( x/\alpha _{\nu }\right) \,\Omega
^{g^{ac_{1}...c_{k}}}\otimes e_{b}\otimes \Omega ^{f^{ac_{1}...c_{k}}}.
\end{array}
\]
In particular, writing ${\sf T}_{a}^{b\;mix}\,\Omega ^{f}=\sum_{g\in
\digamma }\,t_{ab}^{fg}\,\Omega ^{g}$ we have the given sequence of numbers $%
c_{1},...,c_{k}\in 
\mathop{\rm Image}%
f$ defines a function $g$ such that $t_{ab}^{fg}\neq 0$, being 
\begin{equation}
\begin{array}{l}
\Omega ^{g}=\Omega ^{g^{a}}\otimes e_{c_{1}}\otimes \Omega
^{g^{ac_{1}}}\otimes e_{c_{2}}\otimes \Omega ^{g^{ac_{1}c_{2}}}\otimes ...
\\ 
\;\;\;\;\;\;\;\;...\otimes \Omega ^{g^{ac_{1}c_{2}...c_{k-1}}}\otimes
e_{c_{k}}\otimes \Omega ^{g^{ac_{1}...c_{k}}}\otimes e_{b}\otimes \Omega
^{f^{ac_{1}c_{2}...c_{k}}}.
\end{array}
\label{omg}
\end{equation}
Note that $\Omega ^{f}$ can be written 
\begin{equation}
\begin{array}{l}
\Omega ^{f}=\Omega ^{g^{a}}\otimes e_{a}\otimes \Omega ^{g^{ac_{1}}}\otimes
e_{c_{1}}\otimes \Omega ^{g^{ac_{1}c_{2}}}\otimes ... \\ 
\;\;\;\;\;\;\;\;\;\;\;...\otimes \Omega ^{g^{ac_{1}c_{2}...c_{k-1}}}\otimes
e_{c_{k-1}}\otimes \Omega ^{g^{ac_{1}c_{2}...c_{k}}}\otimes e_{c_{k}}\otimes
\Omega ^{f^{ac_{1}c_{2}...c_{k}}}.
\end{array}
\label{omf}
\end{equation}
Furthermore, defining $J_{fg}=\left\{ \sigma _{a},\sigma
_{ac_{1}},...,\sigma _{ac_{1}...c_{k}}\right\} $ and $c_{0}=a$, and
recalling $c_{k+1}=b$, we have [compare with Eq. $\left( \ref{s}\right) $] 
\begin{equation}
\begin{array}{l}
t_{ab}^{fg}=\tau _{b}\,d\,%
\mathop{\textstyle\prod}%
_{j=1}^{k+1}D_{c_{j-1},c_{j}}\,%
\mathop{\textstyle\prod}%
_{\nu \in J_{fg}}G\left( x/\alpha _{\nu }\right) .
\end{array}
\label{tab}
\end{equation}
If $\#\left[ f^{-1}\left( a\right) \right] =k+1$ and $a=b$, the sequence of
numbers $c_{i}=a$, $i=1,...,k$, corresponds to the vector $\Omega
^{g}=\Omega ^{f}$. Also, $J_{ff}=f^{-1}\left( a\right) $ and accordingly,
since $D_{a,a}=1$, 
\begin{equation}
\begin{array}{l}
t_{aa}^{ff}=\tau _{a}\,d\,\prod_{\nu \in f^{-1}\left( a\right) }G\left(
x/\alpha _{\nu }\right) .
\end{array}
\label{da}
\end{equation}

Comparing $\left( \ref{omg}\right) $ and $\left( \ref{omf}\right) $, we see
that functions $g$ such that $t_{ab}^{fg}\neq 0$ necessarily satisfy 
\begin{equation}
\mathop{\rm Image}%
g\cup \left\{ a\right\} =%
\mathop{\rm Image}%
f\cup \left\{ b\right\} .  \label{L}
\end{equation}
In addition, for each element $\mu \in 
\mathop{\rm Image}%
f$, $\mu \neq a,b$, function $g$ must hold 
\begin{equation}
\#\left[ g^{-1}\left( \mu \right) \right] =\#\left[ f^{-1}\left( \mu \right) %
\right] ,  \label{LLd}
\end{equation}
and 
\begin{equation}
\begin{array}{l}
\#\left[ g^{-1}\left( a\right) \right] =\#\left[ f^{-1}\left( a\right) 
\right] -\left( 1-\delta _{ab}\right) , \\ 
\#\left[ g^{-1}\left( b\right) \right] =\#\left[ f^{-1}\left( b\right) 
\right] +\left( 1-\delta _{ab}\right) .
\end{array}
\label{LLLd}
\end{equation}
Now, defining the classes of functions $C\subset \digamma $, in such a way
that $f,g\in C$ {\em iff }

\[
\mathop{\rm Image}%
f=%
\mathop{\rm Image}%
g\;\;\;\;and\;\;\;\;\#\left[ f^{-1}\left( \mu \right) \right] =\#\left[
g^{-1}\left( \mu \right) \right] 
\]
for all $\mu $ contained in their respective images, we can write 
\begin{equation}
\begin{array}{l}
{\sf T}_{a}^{b\;mix}\,\Omega ^{f}=\sum_{g\in
C_{-a}^{+b}}\,t_{ab}^{fg}\,\Omega ^{g},\;\;\;if\;\;f\in C,
\end{array}
\label{mast}
\end{equation}
where $C_{-a}^{+b}$ is the class given by functions $g$ satisfying $\left( 
\ref{L}\right) $, $\left( \ref{LLd}\right) $ and $\left( \ref{LLLd}\right) $%
. Let us note that $C=C_{-a}^{+b}$ {\em iff} $a=b$. Then, denoting 
\begin{equation}
\begin{array}{l}
{\cal H}_{X}\doteq span\left\{ \Omega ^{f}\in {\cal H}^{mix}:f\in X\right\}
\;\;for\;each\;X\subset \digamma ,
\end{array}
\label{dec}
\end{equation}
spaces ${\cal H}_{C}$ are ${\sf T}_{a}^{a\;mix}$-invariant for all $a$. On
the other hand, when $a\neq b$ (since $C\neq C_{-a}^{+b}$), vector $\Omega
^{f}$ can not be written as a linear combination of vectors $\Omega ^{g}$'s
appearing in $\left( \ref{mast}\right) $ (they form a linearly independent
set of vectors). That is, ${\sf T}_{a}^{b\;mix}\,\Omega ^{f}$ is not
proportional to $\Omega ^{f}$. Also, if $a\notin 
\mathop{\rm Image}%
f$, then $C_{-a}^{+b}=\emptyset $ and consequently ${\sf T}%
_{a}^{b\;mix}\,\Omega ^{f}=0$, such as follows from Eq. $\left( \ref{adb}%
\right) $ for $a\neq b$. Last observations translate for operators ${\sf B}%
_{j}^{mix}$ into equations 
\begin{equation}
{\sf B}_{j}^{mix}\,\Omega ^{f}\neq c\,\Omega ^{f},\;\;if\;f^{-1}\left(
1\right) \neq \emptyset ;\;\;{\sf B}_{j}^{mix}\,\Omega ^{f}=0\;\;\;otherwise.
\label{eqII}
\end{equation}

\bigskip

Let us briefly study the reducibility of the action on ${\cal H}^{mix}$ of
the algebra generated by operators ${\sf T}^{mix}$. It follows from Eq. $%
\left( \ref{ii}\right) $ that, if $M=max_{\nu }\left\{ 
\mathop{\rm rank}%
\zeta _{\nu }\right\} $, numbers $c_{1},...,c_{k}$ and $c_{k+1}=b$ must be
smaller than or equal to $M$. This implies ${\sf T}_{a}^{b\;mix}=0$ for $b>M$%
, and we can restrict ourself to the $a,b\leq M$ case. Also, comparing $%
\left( \ref{omg}\right) $ and $\left( \ref{omf}\right) $, if $f\left( \nu
\right) >M$ then $g\left( \nu \right) =f\left( \nu \right) $. As a
consequence, beside $\left( \ref{L}\right) $, $\left( \ref{LLd}\right) $ and 
$\left( \ref{LLLd}\right) $, condition 
\begin{equation}
g\left( \nu \right) =f\left( \nu \right) \;\;\forall \nu \in {\Bbb Z}%
_{N}\;\;such\;that\;\;g\left( \nu \right) ,f\left( \nu \right) >M  \label{l}
\end{equation}
is necessary in order to have $t_{ab}^{fg}\neq 0$. Thus, defining the
classes $E\subset \digamma $ as those whose functions satisfy $\left( \ref{l}%
\right) $, it is clear that spaces ${\cal H}_{E}$ are invariant under the
action of ${\sf T}^{mix}$. It actually can be found smaller invariant spaces
inside ${\cal H}_{E}$, depending {\em locally} on the ranks of glueing
matrices, but we will not discuss it here.

For the equal rank case we have $m=M$, and accordingly the classes $E$ are
in bijection with elements of $\digamma _{0}$. Thus, we can decompose ${\cal %
H}^{mix}$ into ${\sf T}^{mix}$-invariant subspaces ${\cal H}_{E\left(
f\right) }$ labeled by elements of $\digamma _{0}$. In addition, by a simple
inspection of coefficients $\left( \ref{tab}\right) $, it can be shown the
actions on ${\cal H}_{E\left( f\right) }$ and ${\cal H}_{E\left( g\right) }$
are equivalent provided $f^{-1}\left( 1\right) =g^{-1}\left( 1\right) $.
Moreover, in the homogeneous case, namely, $\alpha _{\nu }=1$ for all $\nu $%
, above equivalence still holds when $\#\left[ f^{-1}\left( 1\right) \right]
=\#\left[ g^{-1}\left( 1\right) \right] $.

\bigskip

In the following subsection we diagonalize (when possible) the operator $%
{\sf A}^{mix}$ restricted to each ${\cal H}_{C}$, and show its eigenvectors,
when $C\subset \digamma _{0}$, are precisely the pseudovacuum vectors we are
looking for.

\subsubsection{Diagonalization of ${\sf A}^{mix}$ and vectors $\Phi ^{f}$}

Let us consider a class of functions $C$. Using Eqs. $\left( \ref{da}\right) 
$ and $\left( \ref{mast}\right) $ for $a=b=1$, and defining $a_{fg}\doteq
t_{11}^{fg}$ for $f\neq g$, we have that 
\begin{equation}
\begin{array}{l}
{\sf A}^{mix}\,\Omega ^{f}=a_{f}\,\Omega ^{f}+\sum_{g\in C,g\neq
f}\,a_{fg}\,\Omega ^{g},\;\;\;\;a_{f}=\tau _{1}\,d\,\prod_{\nu \in
f^{-1}\left( 1\right) }G\left( x/\alpha _{\nu }\right) .
\end{array}
\label{masta}
\end{equation}
Now, we are going to show there exists a total order relation between the
functions of $C$, such that w.r.t. this order we can write 
\begin{equation}
\begin{array}{l}
{\sf A}^{mix}\,\Omega ^{f}=a_{f}\,\Omega ^{f}+\sum_{g<f}\,a_{fg}\,\Omega
^{g}.
\end{array}
\label{tri}
\end{equation}
In other words, operator ${\sf A}^{mix}$ restricted ${\cal H}_{C}$ is
represented by a triangular matrix w.r.t. the resulting ordered basis
[recall Eq. $\left( \ref{dec}\right) $ for $X=C$].

To see that, let us consider a function $f\in $ $C$. Assign to $f_{\nu }$
the number $1$ if $f_{\nu }=1$ or $0$ if $f_{\nu }\neq 1$. Denote $b^{f}$
the binary expression related to the sequence $f_{0},...,f_{N-1}$. From Eqs. 
$\left( \ref{omg}\right) $ and $\left( \ref{omf}\right) $ for $a=b=1$, we
see that $b^{f}>b^{g}$ (as real numbers) for $f\neq g$, since unless one $%
f_{\nu }=1$ were moved to the right. This implies $a_{fg}=0$ if $b^{f}\leq
b^{g}$, that is, 
\begin{equation}
\begin{array}{l}
{\sf A}^{mix}\,\Omega ^{f}=a_{f}\,\Omega
^{f}+\sum_{b^{g}<b^{f}}\,a_{fg}\,\Omega ^{g}.
\end{array}
\label{du}
\end{equation}
So let us define an order $<$ between the elements of $C$ by saying $g<f$ if 
$b^{g}<b^{f}$, and when $b^{g}=b^{f}$ we choose an arbitrary order. Using
that and equation above, Eq. $\left( \ref{tri}\right) $ follows immediately.

Since eigenvalues of ${\sf A}^{mix}$ are given by the numbers $a_{f}$, in
order to insure its diagonalizability we can ask the considered model to be
completely inhomogeneous, i.e., $\alpha _{\nu }\neq \alpha _{\mu }$ for all $%
\nu ,\mu \in {\Bbb Z}_{N}$. Then $a_{f}\neq a_{g}$ provided $f^{-1}\left(
1\right) \neq g^{-1}\left( 1\right) $. Thus, eigenvalues are distinct,
unless those related to $f$-th and $g$-th rows for which $f^{-1}\left(
1\right) =g^{-1}\left( 1\right) $. But $f^{-1}\left( 1\right) \neq
g^{-1}\left( 1\right) $ if $b^{f}>b^{g}$. Therefore [see Eq. $\left( \ref{du}%
\right) $], ${\sf A}^{mix}$ does not mix vectors related to rows with the
same diagonal entries, and accordingly ${\sf A}^{mix}$ is diagonalizable.
Actually, we just can insure ${\sf A}^{mix}={\sf A}^{mix}\left( x\right) $
is diagonalizable for almost all values $x$ of the spectral parameter. Note
that for some isolated points $x_{o}\in {\Bbb {C}}$, we can have $%
a_{f}\left( x_{o}\right) =a_{g}\left( x_{o}\right) $, in spite of condition $%
f^{-1}\left( 1\right) \neq g^{-1}\left( 1\right) $ holds.

Using usual recursion formulae for diagonalizing triangular matrices, we can
define for each subspace ${\cal H}_{C}$ the basis $\Phi ^{f}$, $f\in C$,
given by 
\begin{equation}
\Phi ^{f}=\left\{ 
\begin{array}{ll}
\Omega ^{min_{C}}, & \;\;f=min_{C}, \\ 
\Omega ^{f}+\sum_{g<f}\chi _{fg}\,\Phi ^{g}, & \;\;f>min_{C},
\end{array}
\right.  \label{of}
\end{equation}
with 
\begin{equation}
\chi _{fg}=\left\{ 
\begin{array}{ll}
\left. a_{g^{+}g}\right/ \left( a_{g^{+}}-a_{g}\right) , & \;\;f=g^{+}, \\ 
\left. \left( a_{fg}\,-\sum_{g<h<f}a_{fh}\,\chi _{hg}\right) \right/ \left(
a_{f}-a_{g}\right) , & \;\;f>g^{+}.
\end{array}
\right.  \label{gi}
\end{equation}
Here $min_{C}$ is the minimal $f\in C$ w.r.t. the defined order, and $g^{+}$
is the first element in $C$ bigger than $g$. Equation $\left( \ref{gi}%
\right) $ must be understood as a recursive formula on $f$ for each $g$.
Because $\left[ {\sf A}^{mix}\left( x\right) ,{\sf A}^{mix}\left( x^{\prime
}\right) \right] =0$ for all $x,x^{\prime }\in {\Bbb C}$ [that follows from
commutation relations given in $\left( \ref{fcr}\right) $], operators ${\sf A%
}^{mix}\left( x\right) $ can be diagonalized simultaneously. Thus numbers $%
\chi _{fg}$ and vectors $\Phi ^{f}$ do not depend on the spectral parameter.

Let us note diagonal entries ${\sf D}_{ii}^{mix}$ can be diagonalized as
above. But this is not enough to diagonalize ${\sf t}^{mix}$, since
operators ${\sf A}^{mix}$ and ${\sf D}_{ii}^{mix}$ do not commute among
themselves. Nevertheless, last operators restricted to ${\cal H}_{0}$ do
commute, and accordingly can be simultaneously diagonalized. This follows
from the facts that ${\cal H}_{0}\subset \ker \noindent \noindent {\sf C}%
_{i}^{mix}$ and that Eq. $\left( \ref{fcr}\right) $ implies 
\[
\,\noindent \noindent \left[ \,{\sf D}_{ii}^{mix}\left( x\right) ,\noindent
\noindent {\sf A}^{mix}\left( y\right) \right] =-{\sf B}_{i}^{mix}\left(
x\right) \,\noindent \noindent {\sf C}_{i}^{mix}\left( y\right) \,\noindent
\noindent c_{-}\left( x/y\right) +{\sf B}_{i}^{mix}\,\noindent \noindent
\left( y\right) \,{\sf C}_{i}^{mix}\left( x\right) \,c_{+}\left( x/y\right)
. 
\]

\bigskip

Now, let us see that vectors $\Phi ^{f}$ for $f\in \digamma _{0}$, given by $%
\left( \ref{of}\right) $ and $\left( \ref{gi}\right) $, satisfy Equations $%
\left( \ref{II}\right) $ and $\left( \ref{III}\right) $. Since they are
eigenvectors of ${\sf A}^{mix}$ with eigenvalues $a_{f}$, the first part of $%
\left( \ref{II}\right) $ follows immediately. For the second part, note $%
\Phi ^{f}$ is a linear combination of vectors $\Omega ^{g}$ with $g$ inside $%
C$. Also note, if $f\in \digamma _{0}$, then the class defined by $f$ is
inside $\digamma _{0}$ too. Hence, using Eq. $\left( \ref{eqI}\right) $ we
arrive at the wanted result. The same happens for $\left( \ref{III}\right) $
using Eq. $\left( \ref{eqII}\right) $.

For the equal rank case, it can be shown that $a_{fg}=0$ for all $f\in
\digamma _{0}$. In fact, sets $I_{1c_{1}...c_{j}}$ defined by $\left( \ref
{ii}\right) $ (putting $a=b=1$) has $1$ as the unique element, and
consequently the only possible sequence is $c_{i}=1$ for $i=1,...,\#\left[
f^{-1}\left( 1\right) \right] -1$. Such sequence correspond to the vector $%
\Omega ^{f}$. Then, the latter is an eigenvector of ${\sf A}^{mix}$ (without
any inhomogeneity condition). In other words, ${\sf A}^{mix}$ restricted to $%
{\cal H}_{0}$ is represented by a diagonal matrix for the basis $\Omega ^{f}$%
, $f\in \digamma _{0}$, and accordingly $\Phi ^{f}=\Omega ^{f}$.

\bigskip

To end this subsection let us say last results, valid for monodromy matrices 
${\sf T}^{mix}$ of the form $\left( \ref{canm}\right) $, also holds for
those given by Eq. $\left( \ref{cang}\right) $ and satisfying condition $%
\left( \ref{co}\right) $. In fact, on the canonical basis $%
e_{1},...,e_{n_{\nu }}$ of ${\Bbb C}^{n_{\nu }}$, using Eq. $\left( \ref{re}%
\right) $ and $\left( \ref{compa}\right) $, we have that 
\[
\widetilde{{\sf R}}_{a}^{b\;n_{\nu }}\,e_{k}=\left\{ 
\begin{array}{ll}
\delta _{a}^{b}\,/G\left( x/\alpha _{\nu }\right) \,e_{k}, & k\neq a, \\ 
\left( \delta _{a}^{b}+\left( 1-\delta _{a}^{b}\right) \,\left( x/\alpha
_{\nu }\right) ^{2\varepsilon _{ab}^{\nu }}\,c_{%
\mathop{\rm sg}%
\left( b-a\right) }\left( x/\alpha _{\nu }\right) \right) \,e_{b}, & k=a.
\end{array}
\right. 
\]
Then, applying $\widetilde{{\sf T}}_{a}^{b\;mix}$ to a vector $\Omega ^{f}$
we arrive at Eqs. $\left( \ref{adb}\right) $ or $\left( \ref{to}\right) $,
depending on $%
\mathop{\rm Image}%
f$, where the second term of coefficients $C_{i}$ [see Eq. $\left( \ref{to}%
\right) $] must be just changed by a factor $\left( x/\alpha _{\sigma
}\right) ^{2\varepsilon _{ai}^{\sigma }}$. Therefore, all above results{\bf %
\ }follows. In particular, all we have said for ${\sf A}^{mix}$ is also true
for $\widetilde{{\sf A}}^{mix}$, and the former is diagonalizable {\em iff }%
so is the latter. There is a minor change in coefficients $a_{fg}$, and
consequently in the linear combinations $\left( \ref{of}\right) $ that
define eigenvectors of $\widetilde{{\sf A}}^{mix}$. Denoting the latter by $%
\widetilde{\Phi }^{f}$ , and recalling Eq. $\left( \ref{tr}\right) $, we
conclude

\bigskip

{\bf Theorem 1. }Given a mixed vertex model ${\sf T}^{mix}=\zeta _{0}\cdot 
{\sf R}_{0}\stackrel{.}{\otimes }\zeta _{1}\cdot {\sf R}_{1}\stackrel{.}{%
\otimes }...\stackrel{.}{\otimes }\zeta _{N-1}\cdot {\sf R}_{N-1}\,\Upsilon $%
, with glueing matrices $\zeta _{\nu }=P_{\nu }\,\widehat{\zeta }_{\nu
}\,P_{\nu }^{\prime }$ satisfying Eq. $\left( \ref{co}\right) $ and $\left( 
\ref{coa}\right) $, and assuming ${\sf A}^{mix}$ is diagonalizable (e.g., $%
{\sf T}^{mix}$ is completely inhomogeneous), it follows that vectors 
\[
\Phi ^{f}\doteq \left( \otimes _{\nu \in {\Bbb Z}_{N}}P_{\nu }^{\prime
}\right) ^{-1}\,\widetilde{\Phi }^{f},\;\;f\in \digamma _{0}, 
\]
are pseudovacuum states for ${\sf T}^{mix}$ satisfying Eqs. $\left( \ref{II}%
\right) $ and $\left( \ref{III}\right) $. When $%
\mathop{\rm rank}%
\zeta _{\nu }=m$ $\forall \nu $, ${\sf A}^{mix}$ is diagonalizable and $\Phi
^{f}\doteq \left( \otimes _{\nu \in {\Bbb Z}_{N}}P_{\nu }^{\prime }\right)
^{-1}\,\Omega ^{f}$.\ \ \ $\;\Box $

\bigskip

All that can be rephrased in terms of our original mixed monodromy matrices,
i.e., in the form $\left( \ref{mvm}\right) $. We just must regard them as
particular cases of $\left( \ref{ca}\right) $ subject to $\left( \ref{lt}%
\right) $.

\subsection{Nested Bethe equations}

Let ${\sf T}^{mix}$ be a monodromy matrix as that given in theorem above.
Thanks to the algebra embeddings ${\rm YB}_{n-1}\hookrightarrow {\rm YB}_{n}$%
, $n>1$, which are a direct consequence of equations 
\[
\left[ R_{n-1}\right] _{ab}^{kl}=\left[ R_{n}\right] _{ab}^{kl},\;\;for\;1%
\leq a,b,k,l\leq n-1,
\]
it follows that ${\sf T}_{a}^{b\;mix}$ for $a,b\leq m$ satisfy relations
corresponding to the YB algebra ${\rm YB}_{m}$. Then, following for each $%
\Phi ^{f}\in {\cal H}_{0}$ analogous technics to the ones developed in refs. 
\cite{dev}\cite{bab}, that is, proposing as eigenstates for ${\sf t}^{mix}$
[see $\left( \ref{tm}\right) $] the Bethe vectors 
\[
\Psi ^{f}=\Psi ^{j_{1}...j_{r_{1}}}\,{\sf B}_{j_{1}}^{mix}\left( x_{1};{\bf %
\alpha }\right) \,\,...\,{\sf B}_{j_{r}}^{mix}\left( x_{r};{\bf \alpha }%
\right) \,\Phi ^{f},\;\;j_{1},...,j_{r}<m,
\]
and separating in the so-called {\em wanted} and {\em unwanted} terms, we
arrive at a set of nested Bethe ansatz equations which in its recursive form
are given by 
\begin{equation}
\begin{array}{c}
{\displaystyle{%
\mathop{\textstyle\prod}_{p=1}^{r_{1}}G\left( x_{k}^{\left( 1\right) }/x_{p}^{\left( 1\right) }\right)  \over %
\mathop{\textstyle\prod}_{\nu \in f^{-1}\left( 1\right) }G\left( x_{k}^{\left( 1\right) }/\alpha _{\nu }\right) }}%
\,\Lambda _{1}\left( x_{k}^{\left( 1\right) }\right) +\tau _{1}\,%
\mathop{\textstyle\prod}%
_{p=1}^{r_{1}}G\left( x_{p}^{\left( 1\right) }/x_{k}^{\left( 1\right)
}\right) =0, \\ 
\\ 
{\displaystyle{%
\mathop{\textstyle\prod}_{p=1}^{r_{l}}G\left( x_{k}^{\left( l\right) }/x_{p}^{\left( l\right) }\right)  \over %
\mathop{\textstyle\prod}_{v=1}^{r_{l-1}}G\left( x_{k}^{\left( l\right) }/x_{v}^{\left( l-1\right) }\right) }}%
\,\Lambda _{l}\left( x_{k}^{\left( l\right) }\right) +\tau _{l}\,%
\mathop{\textstyle\prod}%
_{p=1}^{r_{l}}G\left( x_{p}^{\left( l\right) }/x_{k}^{\left( l\right)
}\right) =0,\;\;\left( l>1\right) 
\end{array}
\label{be}
\end{equation}
and 
\begin{equation}
\begin{array}{c}
\Lambda _{m-1}\left( x\right) =\tau _{m}, \\ 
\Lambda _{l}\left( x\right) =\,%
{\displaystyle{%
\mathop{\textstyle\prod}_{p=1}^{r_{l+1}}G\left( x/x_{p}^{\left( l+1\right) }\right)  \over %
\mathop{\textstyle\prod}_{u=1}^{r_{l}}G\left( x/x_{u}^{\left( l\right) }\right) }}%
\,\Lambda _{l+1}\left( x\right) +\tau _{l+1}\,%
\mathop{\textstyle\prod}%
_{p=1}^{r_{l+1}}G\left( x_{p}^{\left( l+1\right) }/x\right) ,\;\;\left(
l<m-2\right) 
\end{array}
\label{wt}
\end{equation}
where $l=1,...,m-1$, $k=1,...,r_{l}$, and $0\leq r_{l}$ $\leq r_{l-1}\leq \#%
\left[ f^{-1}\left( 1\right) \right] $. Thus, Bethe equations related to a
given $\Phi ^{f}\in {\cal H}_{0}$, are the ones corresponding to an $A_{m-1}$
type quasi-periodic vertex model with $n_{f}\doteq \#\left[ f^{-1}\left(
1\right) \right] $ sites per row and inhomogeneity vector ${\bf \alpha }%
_{f}=\left( \alpha _{\nu _{0}},\alpha _{\nu _{1}},...,\alpha _{\nu
_{n_{f}-1}}\right) $, such that $\nu _{i}\in f^{-1}\left( 1\right) $ and $%
\nu _{i}<\nu _{i+1}$ for all $i\in {\Bbb Z}_{n_{f}}$. When $\#\left[
f^{-1}\left( 1\right) \right] =0$ we have no Bethe equations. Note in this
case ${\sf B}_{j}^{mix}\,\Phi ^{f}=0$ $\forall j$ [see Eq. $\left( \ref{III}%
\right) $]. For each solution 
\[
{\bf x}=\left\{ {\bf x}^{\left( l\right) }=\left( x_{1}^{\left( l\right)
},...,x_{r_{l}}^{\left( l\right) }\right) :l=1,...,m-1\right\} 
\]
of Equations $\left( \ref{be}\right) $ and $\left( \ref{wt}\right) $,

\begin{equation}
\begin{array}{c}
\Lambda ^{f}\left( x;{\bf x}\right) =d\,%
\mathop{\textstyle\prod}%
_{k=1}^{r_{1}}G\left( x/x_{k}^{\left( 1\right) }\right) \,\Lambda _{1}\left(
x\right) +\tau _{1}\,d\,\prod_{\nu \in f^{-1}\left( 1\right) }G\left(
x/\alpha _{\nu }\right) \,%
\mathop{\textstyle\prod}%
_{k=1}^{r_{1}}G\left( x_{k}^{\left( 1\right) }/x\right)
\end{array}
\label{lamb}
\end{equation}
gives an eigenvalue of ${\sf t}^{mix}$. Note that $\Lambda ^{f}\left( x;{\bf %
x}\right) =\Lambda ^{g}\left( x;{\bf x}\right) $ if $f^{-1}\left( 1\right)
=g^{-1}\left( 1\right) $. This is the main source of degeneracy for the
transfer matrix. It can be seen each $\Lambda ^{f}\left( x;{\bf x}\right) $
differs by a factor $\prod_{\nu \notin f^{-1}\left( 1\right) }1/G\left(
x/\alpha _{\nu }\right) $ from the corresponding eigenvalue related to the
mentioned $A_{m-1}$ model. Eigenvectors $\Psi ^{f}\left( {\bf x}\right) $,
i.e., the Bethe vectors, can also be given recursively, but now through
vectors $\Psi _{l}\in \left( {\Bbb C}^{m-l}\right) ^{\otimes r_{l}}$ with
coordinates $\left( \Psi _{l}\right) ^{j_{1}...j_{r_{l}}}$ (w.r.t. the
canonical basis of ${\Bbb C}^{m-l}$) such that 
\begin{equation}
\Psi ^{f}\left( {\bf x}\right) =\left( \Psi _{1}\right)
^{j_{1}...j_{r_{1}}}\,{\sf B}_{j_{1}}^{mix}\left( x_{1}^{\left( 1\right) };%
{\bf \alpha }\right) \,\,...\,{\sf B}_{j_{r_{1}}}^{mix}\left(
x_{r_{1}}^{\left( 1\right) };{\bf \alpha }\right) \,\Phi ^{f},  \label{bv}
\end{equation}
for $1\leq l\leq m-2$ 
\[
\Psi _{l}=\left( \Psi _{l+1}\right) ^{j_{1}...j_{r_{l+1}}}\,{\sf B}%
_{j_{1}}^{\left( m-l,r_{l}\right) }\left( x_{1}^{\left( l+1\right) };{\bf x}%
^{\left( l\right) }\right) \,\,...\,{\sf B}_{j_{r_{l+1}}}^{\left(
m-l,r_{l}\right) }\left( x_{r_{l+1}}^{\left( l+1\right) };{\bf x}^{\left(
l\right) }\right) \,\omega _{l}, 
\]
and $\Psi _{m-1}=1$. Here $j_{1},...,j_{r_{l+1}}<m$. We are denoting by $%
\omega _{l}$ the pseudovacuum for the pure monodromy matrix ${\sf T}^{\left(
m-l,r_{l}\right) }$. Let us mention, in the $l$-th level of nesting process
the involved monodromy matrix actually is the twisting 
\[
{\sf T}^{\left( m-l,r_{l}\right) }\cdot \Upsilon
_{l},\;\;\;being\;\;\Upsilon _{l}=diag\left( \tau _{1},...,\tau
_{m-l}\right) , 
\]
which also has $\omega _{l}$ as pseudovacuum vector.

Summing up, we have constructed a set of eigenvectors for ${\sf t}^{mix}$ by
applying creation operators ${\sf B}_{j}^{mix}$'s over all $\Phi ^{f}$, $%
f\in \digamma _{0}$. In the following section we address the combinatorial
completeness of that set of states.

\bigskip

By last, let us say that equations $\left( \ref{be}\right) $ and $\left( \ref
{wt}\right) $ do not depend neither on permutations $P_{\nu },P_{\nu
}^{\prime }$ defining the glueing matrices of ${\sf T}^{mix}$ (recall
conditions of theorem above), nor on the set of ranks of the latter. They
only depends on the minimum $m=min_{\nu }\left\{ 
\mathop{\rm rank}%
\zeta _{\nu }\right\} $ of that set, on the boundary matrix $\Upsilon $, and
on the inhomogeneity vector ${\bf \alpha }$. Hence, assuming complete
integrability, the spectrum of the related transfer matrix ${\sf t}^{mix}$,
which would be given by the numbers $\Lambda ^{f}\left( x;{\bf x}\right) $
defined in $\left( \ref{lamb}\right) $, only depends on $m$, $\Upsilon $ and 
${\bf \alpha }$. Accordingly,

\bigskip

{\bf Theorem 2.} Assuming complete integrability, every mixed model with
glueing matrices satisfying Equation $\left( \ref{co}\right) $ is physically
equivalent to one with monodromy matrix of the form $\left( \ref{canm}%
\right) $ and satisfying the equal rank condition: $%
\mathop{\rm rank}%
\zeta _{\nu }=m$ for all $\nu $.$\;\;\;\Box $

\subsection{Combinatorial completeness}

In this section we are going to show that Eq. $\left( \ref{bv}\right) $
(varying indices $j$ from $1$ to $m$, functions $f$ in $\digamma _{0}$, and $%
{\bf x}$ along solutions of $\left( \ref{be}\right) $ and $\left( \ref{wt}%
\right) $) defines unless $\dim {\cal H}^{mix}=\prod_{\nu \in {\Bbb Z}%
_{N}}n_{\nu }$ different vectors. That is to say, we have a set of Bethe
vectors from which, {\em a priori}, a basis of eigenstates for the related
transfer matrix can be extracted. To see that, we shall assume combinatorial
completeness of Bethe ansatz equations related to the $A_{n-1}$ vertex
models, i.e., for a model with $N$ sites in a row we suppose there is unless 
$\left( n-1\right) ^{r}\,%
{N \choose r}%
$ different solutions for the Bethe equations corresponding to $r$ creation
operators. This has been shown for $n=2$ (see for instance \cite{kir}), but
we do not know about any similar result for bigger $n$. In our case, we
would be saying for each vector $\Phi ^{f}$ with $f\in \digamma _{0}$, there
exists unless a number $\left( m-1\right) ^{r}\,%
{n_{f} \choose r}%
$ of different solutions of $\left( \ref{be}\right) $ and $\left( \ref{wt}%
\right) $ corresponding to $r_{1}=r$ creation operators. Recall that $%
n_{f}=\#\left[ f^{-1}\left( 1\right) \right] $. Let us first see why this
assumption is useful for our purposes.

\bigskip

It is enough to analyze the case of monodromy matrices given by $\left( \ref
{canm}\right) $. The other cases, i.e., those given by $\left( \ref{can}%
\right) $ and satisfying $\left( \ref{co}\right) $, follow analogously. So
let us come back to \S {\bf IV.A.1} and consider the action of operators $%
{\sf B}_{j}^{mix}$ with $j<m$, on vectors $\Omega ^{f}$ with $f\in \digamma
_{0}$. Suppose first that $%
\mathop{\rm rank}%
\zeta _{\nu }=m$ for all $\nu $. For $a=1$ and $b=j+1$, sequences $c_{i}=1$, 
$i=1,...,k$, with $1\leq k<n_{f}$ define terms proportional to vectors $%
\Omega ^{g}=\Omega ^{f_{\mu ,j}}$, with $\mu \in f^{-1}\left( 1\right) $, $%
f_{\mu ,j}\left( \nu \right) =f\left( \nu \right) $ for all $\nu \neq \mu $
and $f_{\mu ,j}\left( \mu \right) =j+1$. That is, we change a vector $e_{1}$
by a vector $e_{j+1}$ in position $\mu \in f^{-1}\left( 1\right) $. They are
the only possible sequences. Thus, the action of each ${\sf B}_{j}^{mix}$, $%
j=1,...,m-1$, on a vector $\Omega ^{f}$ gives rise to a linear combination
of $n_{f}$ linearly independent vectors. Existence of $n_{f}$ different
solutions to Eqs. $\left( \ref{be}\right) $ and $\left( \ref{wt}\right) $
for $r_{1}=1$ and for each $j$, is a necessary condition to obtain $n_{f}$
l.i. eigenstates from the set of Bethe vectors. Then, varying $j$ from $1$
to $m-1$, we shall have, {\em a priori}, $\left( m-1\right) \,n_{f}$ l.i.
eigenstates. Applying ${\sf B}_{i}^{mix}$ and ${\sf B}_{j}^{mix}$ we have $%
\left( m-1\right) ^{2}\,$vectors, each one of them having $n_{f}\,\left(
n_{f}-1\right) /2$ l.i. terms. In general, if we apply $r$ creation
operators to $\Omega ^{f}$, we have $\left( m-1\right) ^{r}$ vectors with
related $\,%
{n_{f} \choose r}%
$ terms. Now it becomes clear why our assumption is needed. The same
argument can be given for the general rank case. There, when an operator $%
{\sf B}_{j}^{mix}$ acts on $\Omega ^{f}$ we have as above the terms
proportional to $\Omega ^{f_{\mu ,j}}$, $\mu \in f^{-1}\left( 1\right) $,
together with additional terms given by vectors $\Omega ^{h_{\sigma ,j}}$
with $h$ belonging to the same class of $f$. Thus, the latter appears as
terms when ${\sf B}_{j}^{mix}$ is applied to $\Omega ^{h}$. Accordingly, in
order to avoid overcounting, we do not have to take them into account.

\bigskip

Let us come back to our original problem. If combinatorial completeness
holds there exists unless a number $\sum_{r=0}^{n_{f}}\left( m-1\right)
^{r}\,%
{n_{f} \choose r}%
=\left( \left( m-1\right) +1\right) ^{n_{f}}=m^{n_{f}}$ of Bethe vectors for
each function $f\in \digamma _{0}$. Thus, since $0\leq n_{f}\leq N$ for
every $f\in \digamma $, the total number of Bethe vectors is $\sum_{f\in
\digamma _{0}}m^{n_{f}}=\sum_{k=0}^{N}m^{k}\,p_{k}$, being $p_{k}$ the
number of functions $f\in \digamma _{0}$ such that $n_{f}=k$. Let us
calculate $p_{k}$. It is clear that the number of functions $f$ in $\digamma
_{0}$ with the same pre-image $f^{-1}\left( 1\right) $ is 
\begin{equation}
\prod_{\nu \in {\Bbb Z}_{N}}\left( n_{\nu }-m\right) ^{\varepsilon _{\nu
}},\;\;\;\varepsilon _{\nu }=\left\{ 
\begin{array}{l}
0,\;\;\nu \in f^{-1}\left( 1\right) , \\ 
1,\;\;otherwise.
\end{array}
\right.  \label{mult}
\end{equation}
In terms of numbers $\varepsilon _{0},...,\varepsilon _{N-1}$, the condition 
$n_{f}=\#\left[ f^{-1}\left( 1\right) \right] =k$ can be characterized by
equality $\varepsilon _{0}+...+\varepsilon _{N-1}=N-k$. Then, in order to
obtain $p_{k}$ we must sum over all configurations of $\varepsilon
_{0},...,\varepsilon _{N-1}$ ($\varepsilon _{\nu }$ equal to $0$ or $1$),
such that last condition holds, i.e., 
\begin{equation}
\begin{array}{l}
p_{k}=\sum_{\varepsilon _{0},...,\varepsilon _{N-1}}\prod_{\nu \in {\Bbb Z}%
_{N}}\left( n_{\nu }-m\right) ^{\varepsilon _{\nu }}\,\delta _{\varepsilon
_{0}+...+\varepsilon _{N-1},N-k}.
\end{array}
\label{pk}
\end{equation}
Accordingly, 
\[
\begin{array}{l}
\sum_{k=0}^{N}m^{k}\,p_{k}=\sum_{k=0}^{N}m^{k}\,\left( \sum_{\varepsilon
_{0},...,\varepsilon _{N-1}}\prod_{\nu \in {\Bbb Z}_{N}}\left( n_{\nu
}-m\right) ^{\varepsilon _{\nu }}\,\delta _{\varepsilon _{0}+...+\varepsilon
_{N-1},N-k}\right) \\ 
=\sum_{\varepsilon _{0},...,\varepsilon _{N-1}}\prod_{\nu \in {\Bbb Z}%
_{N}}\left( n_{\nu }-m\right) ^{\varepsilon _{\nu
}}\,\sum_{k=0}^{N}m^{k}\,\delta _{\varepsilon _{0}+...+\varepsilon
_{N-1},N-k} \\ 
=\sum_{\varepsilon _{0},...,\varepsilon _{N-1}}\prod_{\nu \in {\Bbb Z}%
_{N}}\left( n_{\nu }-m\right) ^{\varepsilon _{\nu }}\,\,m^{N-\left(
\varepsilon _{0}+...+\varepsilon _{N-1}\right) } \\ 
=m^{N}\,\sum_{\varepsilon _{0},...,\varepsilon _{N-1}}\prod_{\nu \in {\Bbb Z}%
_{N}}\left( \frac{n_{\nu }}{m}-1\right) ^{\varepsilon _{\nu }}\,\,.
\end{array}
\]
But 
\[
\begin{array}{l}
\sum_{\varepsilon _{0},...,\varepsilon _{N-1}}\prod_{\nu \in {\Bbb Z}%
_{N}}\left( \frac{n_{\nu }}{m}-1\right) ^{\varepsilon _{\nu }}=\prod_{\nu
\in {\Bbb Z}_{N}}\left( \sum_{\varepsilon _{\nu }}\left( \frac{n_{\nu }}{m}%
-1\right) ^{\varepsilon _{\nu }}\right) \\ 
=\prod_{\nu \in {\Bbb Z}_{N}}\left( \left( \frac{n_{\nu }}{m}-1\right)
^{0}+\left( \frac{n_{\nu }}{m}-1\right) \right) =\prod_{\nu \in {\Bbb Z}_{N}}%
\frac{n_{\nu }}{m}=m^{-N}\,\prod_{\nu \in {\Bbb Z}_{N}}n_{\nu },
\end{array}
\]
and consequently $\sum_{f\in \digamma _{0}}m^{n_{f}}=\prod_{\nu \in {\Bbb Z}%
_{N}}n_{\nu }$, as we wanted to see.

\section*{Conclusions}

From last equation we see that, under conditions of {\bf Theor. 1} and
assuming complete integrability, ${\cal H}^{mix}$ can be decomposed into a
direct sum of $m^{n_{f}}$-dimensional spaces ${\cal H}_{f}$, each one of
them generated by the Bethe vectors related with some $f$ inside $\digamma
_{0}$. Note this sum, in general, is not orthogonal w.r.t. the usual scalar
product in $\otimes _{\nu \in {\Bbb Z}_{N}}{\Bbb C}^{n_{\nu }}$. Thinking of
the quantum spin ring related to our vertex model, whose Hamiltonian $H$ is
constructed from the logarithmic derivative (if there exists) of the
transfer matrix, states of ${\cal H}_{f}$ can be interpreted as those of an
anisotropic $A_{m-1}$ type spin chain with $n_{f}$ sites, which are {\em %
localized} on the subring ${\Bbb Z}_{n_{f}}\backsim f^{-1}\left( 1\right)
\subset {\Bbb Z}_{N}$. In other words, we have decomposed a mixed spin model
as a direct sum of $A_{m-1}$ type ones with different numbers of sites and
generically{\em \ }different inhomogeneities. Multiplicity of these models
is given by $\left( \ref{mult}\right) $ [recall eigenvalues $\left( \ref
{lamb}\right) $ only depend on $f$ through $f^{-1}\left( 1\right) $]. In
connection with {\bf Theor. 2} let us say that for the equal rank case,
since we have ${\cal H}_{f}={\cal H}_{E\left( f\right) }$ (see at the end of
\S {\bf IV.A.1}), described decomposition (which results orthogonal) and
mentioned multiplicity are direct consequences of the facts that last spaces
are ${\sf T}^{mix}$-invariant, and that corresponding actions on spaces $%
{\cal H}_{f}$ and ${\cal H}_{g}$ are equivalent when $f^{-1}\left( 1\right)
=g^{-1}\left( 1\right) $.

In the homogeneous case we have in addition actions on ${\cal H}_{f}$ and $%
{\cal H}_{g}$ are equivalent still when $n_{f}=n_{g}$. In other terms, for
the homogeneous equal rank case we can write ${\cal H}^{mix}$ as the
orthogonal direct sum ${\cal H}^{mix}=\bigoplus_{k=0}^{N}{\Bbb C}%
^{p_{k}}\otimes {\cal H}_{f_{k}}$ [see $\left( \ref{pk}\right) $ for numbers 
$p_{k}$], being $f_{k}$ some function with $n_{f_{k}}=k$.

Concluding, we have presented a procedure for glueing different integrable
vertex models in such a way that the integrability of the whole system is
preserved. This procedure relies on some generalization of the coalgebra
structure to the case of rectangular quantum matrices and their
representations, enhancing the deep linking between these algebraic
structures and integrability.

\subsection*{Acknowledgments}

H.M. thanks to CONICET and S.G. thanks to CNEA and Fundaci\'{o}n Antorchas,
Argentina, for financial support.


\begin{references}
\bibitem{baxter}  R. J. Baxter, {\em Exactly solved models in statistical
mechanic} (Academic Press, London, 1982).

\bibitem{faddeev}  L. D. Faddeev, {\em Recent Advances in Field Theory and
Statistical Mechanics, Les Houches, Sesion XXXIX},{\em \ }edited by J -B.
Zuber and R. Stora (North-Holland, Amsterdam, 1982), 561.

\bibitem{d}  V. G. Drinfeld, Proc. Int. Congr. Math. Berkeley {\bf 1}, 798
(1986); M. Jimbo, Lett. Math. Phys. {\bf 11}, 247 (1986).

\bibitem{frt}  N. Yu. Reshetikhin, L. A. Takhtadjian and L. D. Faddeev,
Leningrad Math. J. {\bf 1}, 193 (1990).

\bibitem{jones}  V. R. F. Jones, Int. J. Mod. Phys. B {\bf 4}, 701 (1990).

\bibitem{mm}  S. Majid and M. Markl, {\em Glueing operation for R-matrices,
quantum groups and link invariants of Hecke type}, Math. Proc. Camb. Phil.
Soc. 118 (1995).

\bibitem{dev}  H. J. De Vega, Int. J. Mod. Phys. A {\bf 4}, 2371 (1989).

\bibitem{fad}  L. D. Faddeev, {\em preprint} {\tt hep-th/9605187}.

\bibitem{marti}  C. G\'{o}mez, M. Ruiz Altaba and G. Sierra, {\em Quantum
groups in two dimensional physics} (Cambridge Univ. Press, Cambridge, 1996).

\bibitem{kor}  V. E. Korepin, N. M. Bogoliubov and A. G. Izergin, {\em %
Quantum inverse scattering method and correlation functions} (Cambridge
Monographs in Math. Phys., Cambridge, 1997).

\bibitem{baxter2}  R. Baxter, {\em preprint} {\tt cond-mat/0111188}.

\bibitem{kr}  P. P. Kulish and N. Yu. Reshetikhin, J. Phys. A {\bf 16}, 591
(1983).

\bibitem{dk}  H. J. De Vega and M. Karowski, Nucl. Phys. B {\bf 280}, 225
(1987).

\bibitem{maj}  S. Majid, {\em Foundations of quantum group theory}
(Cambridge University Press, Cambridge, 1995).

\bibitem{hugo}  S. Grillo and H. Montani, {\em in preparation}.

\bibitem{zn}  O. Babelon, H. J. De Vega and C. M. Viallet, Nucl. Phys. B 
{\bf 190}, 542 (1981).

\bibitem{man}  E. E. Demindov, Yu. I. Manin, E. E. Mukhin and D. V.
Zhdanovich, Prog. Theor. Phys. Suppl. {\bf 102}, 203 (1990).

\bibitem{drin}  V. G. Drinfeld, Leningrad Math J. {\bf 1}, 1419 (1990).

\bibitem{foer}  A. Foerster, I. Roditi and L. Rodriguez, Mod. Phys. Lett. A 
{\bf 11}, 987 (1996).

\bibitem{bab}  O. Babelon, H. J. De Vega and C. M. Viallet, Nucl. Phys. B 
{\bf 200}, 266 (1982).

\bibitem{kir}  A. N. Kirilov, J. Soviet Mathematics {\bf 30}, 2298 (1985);
A. N. Kirilov and N. A. Liskova, J. Phys. A {\bf 30}, 1209 (1997).
\end{references}
\end{document}